\documentclass{aastex62}
\hypersetup{linkcolor=red,citecolor=blue,filecolor=cyan,urlcolor=magenta}
\usepackage{rotating}
\usepackage{multirow}
\usepackage{lineno}

\newcommand{\kms}{\mbox{km\,s$^{-1}$}}

\newcommand{\gcm}{\mbox{g cm$^{-2}$}} 
\newcommand{\mjypbm}{\mbox{mJy\,beam$^{-1}$}}
\newcommand{\mujypbm}{\mbox{$\mu$Jy\,beam$^{-1}$}}

\newcommand{\msun}{\mbox{\,$M_\odot$}}
\newcommand{\msunyr}{\mbox{\,$M_\odot$ yr$^{-1}$}}

\graphicspath{{./}{figures/}}

\received{}
\revised{}
\accepted{}
\submitjournal{ApJ}

\shorttitle{The disk population in a distant massive protocluster}
\shortauthors{Cheng et al.}

\begin{document}

\title{The Disk Population in a Distant Massive Protocluster}

\correspondingauthor{Yu Cheng}
\email{ycheng.astro@gmail.com}

\author[0000-0002-8691-4588]{Yu Cheng}
\affil{National Astronomical Observatory of Japan, 2-21-1 Osawa, Mitaka, Tokyo, 181-8588, Japan}

\author[0000-0002-3389-9142]{Jonathan C. Tan} 
\affiliation{Dept. of Space, Earth \& Environment, Chalmers University of Technology, 412 93 Gothenburg, Sweden}
\affiliation{Dept. of Astronomy, University of Virginia, Charlottesville, Virginia 22904, USA}

\author[0000-0002-6195-0152]{John J. Tobin}
\affiliation{National Radio Astronomy Observatory, 520 Edgemont Rd., Charlottesville, VA 22903, USA}

\author[0000-0003-4040-4934]{Rub\'{e}n Fedriani}
\affil{Dept. of Space, Earth \& Environment, Chalmers University of Technology, 412 93 Gothenburg, Sweden}

\author[0000-0002-5306-4089]{Morten Andersen}
\affil{European Southern Observatory, Karl Schwarzschild Str. 2, 85748, Garching, Germany}

\author[0000-0003-4874-0369]{Junfeng Wang}
\affiliation{Department of Astronomy, Xiamen University, Xiamen, Fujian 361005, China}




\begin{abstract}
The unprecedented angular resolution and sensitivity of ALMA makes it possible to unveil disk populations in distant ($>$2 kpc), embedded young cluster environments. We have conducted an observation towards the central region of the massive protocluster G286.21+0.16 at 1.3~mm. With a spatial resolution of 23 mas and a sensitivity of 15~\mujypbm, we detect a total of 38 protostellar disks. These disks have dust masses ranging from about 53 to 1825 $M_\oplus$, assuming a dust temperature of 20~K. 
This sample is not closely associated with previously identified dense cores, as would be expected for disks around Class 0 protostars. Thus, we expect our sample, being flux limited, to be mainly composed of Class I/flat-spectrum source disks, since these are typically more massive than Class II disks.
Furthermore, we find that the distributions of disk masses and radii are statistically indistinguishable with those of the Class I/flat-spectrum objects in the Orion molecular cloud, indicating similar processes are operating in G286.21+0.16 to regulate disk formation and evolution. 
The cluster center appears to host a massive protostellar system composed of three sources within 1200~au, including a potential binary with 600~au projected separation.
Relative to this center, there is no evidence for widespread mass segregation in the disk population. We do find a tentative trend of increasing disk radius versus distance from the cluster center, which may point to the influence of dynamical 
interactions being stronger in the central regions.
\end{abstract}


\section{Introduction}\label{sec:intro}

Circumstellar disks of dust and gas are a common feature of young and forming stellar systems, from deeply embedded Class 0 protostars to more evolved Class II phase or pre-main-sequence stars, at which point the natal envelope has dissipated. These disks play important roles in both star and planet formation \citep[e.g.,][]{Armitage11,Williams11}. Large samples of disks with high-fidelity imaging obtained by interferometers, such as the Atacama Large Millimeter/submillimeter Array (ALMA), have greatly boosted the study of their formation and evolution. In particular, high resolution observations in millimeter continuum allow for constraints on the masses and sizes of dust disks. Surveys of Class II disks to date span a range of physical conditions and ages, e.g., low-mass star-forming regions like Lupus \citep[1--3~Myr,][]{Ansdell16}, Taurus \citep[1--3~Myr,][]{Tripathi17} and Chamaeleon~I \citep[2--3~Myr,][]{Pascucci16}, and more clustered environments, such as the Orion Nebula Cluster \citep[ONC, $<$1~Myr,][]{Eisner18,Otter21}, Orion A \citep[1--3~Myr,][]{VanTerwisga22}, $\sigma$~Orionis \citep[3--5~Myr,][]{Ansdell17}, and the Upper~Scorpius~OB~association \citep[5--10~Myr,][]{Barenfeld16}. 

Meanwhile, the ensemble properties of dust disks in the protostellar phase (Class 0, Class I, and flat-spectrum) are also being established. In the VANDAM Orion survey towards a sample of 328 protostars, \citet{Tobin20} found that masses and radii decreased during the protostellar phase. Still, these protostellar dust disks have masses that are systematically larger than those of Class II disks. Surveys of young disks in Perseus reveal a similar mass distribution of Class I disks as Orion, but more massive Class 0 disks \citep{Tychoniec20}. A systematically less massive and smaller disk population is reported for Ophiuchus \citep{Cieza19,Encalada21}, but this could result from misidentification of disk class, rather than being a real difference \citep{Tobin20}. These observed properties of disks throughout the protostellar phase inform us of both the conditions of their formation and the initial conditions for disk evolution. It is during the protostellar phase that the largest mass reservoir is available for the formation of companions or planets.

Still, further surveys of more diverse samples, especially of massive protoclusters, are necessary to unveil the protostellar disk properties of  environments in which most Galactic star formation occurs, and to explore how they might be affected by external environmental factors \citep[e.g.,][]{Eisner18}. However, the above studies are all limited to relatively nearby regions ($d\lesssim$0.5~kpc). For more distant regions, a complete identification and classification of disks and associated host stars is generally unavailable due to limited spatial resolution and sensitivity for infrared facilities, such as {\it Spitzer Space Telescope} and {\it Herschel Space Observatory}. 
Recently, \citet{Busquet19} showed that it is possible to characterize the disk population in a relatively distant massive protocluster, GGD 27 ($\sim$1.4~kpc), with the long baseline capability of ALMA. In a single pointing they detected 25 compact continuum sources with a resolution of 40~mas ($\sim$56~au) and a sensitivity of mm flux equivalent to a gas mass $\sim$0.002~\msun{} (or a dust mass of $\sim 7 M_\oplus$ assuming a gas to dust ratio of 100). 

Here we present an ALMA long baseline survey of protostellar disks in the central region of a more distant massive protocluster.
Our target G286.21+0.17 (hereafter G286) is a massive, gas-dominated
protocluster associated with the $\eta$ Car giant molecular cloud at a
distance of 2.5$\pm$0.3~kpc \citep{Barnes10}. This distance is also consistent with the recent estimation based on Gaia DR2 parallax measurements \citep{Zucker20}. The gas and dust
components in the wider region covering the whole protocluster have been previously studied with relatively low-resolution ALMA observations, which revealed $\sim80$ dense cores in millimeter continuum emission and molecular line emission \citep{Cheng18,Cheng20a}. In addition,
near-IR (NIR) observations with the VLT found that a high fraction of the young stellar objects (YSOs) have disks, further suggesting
the cluster is very young \citep[$\sim$ 1~Myr;][]{Andersen17}. Follow up multi-epoch HST observations showed that these YSOs with disk excesses also exhibit a higher 
variability fraction and some high amplitude variables are likely associated with accretion outburst events \citep{Cheng20b}. Here we present high-resolution ALMA observations of the central region of G286, which enable detection and study of the disk population.
The paper is organized as follows: a description of the observations
is given in \autoref{sec:obs}; the results are presented in \autoref{sec:res}; a discussion is made in 
\autoref{sec:dis}; and a summary is given in \autoref{sec:sum}.

\section{Observations}\label{sec:obs}

\subsection{ALMA long-baseline observations}

G286 was observed with ALMA (Project ID 2017.1.01552.S, PI: Cheng, Y.) in a single pointing centered on J2000 R.A.=10:38:32.2, Decl.=$-$58:19:8.5, i.e., the position of the
most luminous dense core in this region \citep{Cheng18}.
We employed the
C36-10 configuration to achieve the highest spatial resolution in band~6.
Five executions were conducted during a period from Oct. 2017 to Nov. 2017 , with 44 -- 49 antennas and covering baselines from 40~m to 16000~m. The precipitable water vapor (PWV) during observations ranged from 0.4~mm to 1.2~mm.


The correlator was configured with the first baseband split into two 234.38~MHz spectral windows with 1920 channels each (0.37~\kms{} velocity resolution) and centered on
H30$\alpha$ and CO(2-1), respectively. The second baseband was set to low spectral resolution continuum mode, 1.875 GHz bandwidth divided into 128 31.25~MHz channels, centered at 234.0~GHz. The third baseband was split into three 58.6~MHz spectral windows centered on $\rm C^{18}O~(2-1)$, $\rm H_2CO(3_{2,1}-2_{2,0})$, and $\rm CH_3OH~(4_{2,2}-3_{1,2})$, respectively. Each spectral window has 960 channels, corresponding to 0.19~\kms{} velocity resolution. The fourth baseband was configured with two 234.38~MHz spectral windows (1920
 channels, 0.39~\kms{} velocity resolution) centered on $\rm SiO(5-4)$ and $\rm SO_2(22_{2,20}-22_{1,21})$. 
This paper will focus on the continuum results, while the line results will be presented in a future work.
 
The raw data were calibrated with the ALMA data reduction pipeline using {\it CASA 5.1.1} \citep{McMullin07}. The continuum visibility data were constructed with all line-free channels. In order to increase the signal-to-noise ratio of the continuum, we performed two rounds of phase-only self-calibration, with the first round using a solution interval that encompassed the length of an entire on-source scan, and the second round using a 6.05~s solution interval, which corresponds to a single integration. We created maps with a Briggs weighting scheme with a robust parameter of 0.5, which yields a resolution of 26~mas$\times$20~mas, i.e., 65~au$\times$50~au, with a position angle (P.A.) of 9.3$^{\circ}$. The resultant root mean square (rms) noise in the 1.3~mm continuum was $\sim$15~\mujypbm. The maximum recoverable scale was $\sim$0\farcs{22}, corresponding to 550~au, so our observations are mainly sensitive to very compact structures. The half power beam width (HPBW) is 25\arcsec{} at observed frequency. The flux accuracy from absolute amplitude calibration is expected to be $\lesssim$10\%.

The fiducial image, i.e., made with robust$=$0.5, achieves a good compromise between sensitivity and spatial resolution. To test the robustness of observational results, we also created continuum images with natural and uniform weighting, respectively. For the natural weighting, we obtained a synthesized beam of 31~mas$\times$25~mas and a rms noise of 15~\mujypbm. For the uniform weighting, we obtained an image with a resolution of 20~mas$\times$14~mas and a rms noise of 34~\mujypbm.

\subsection{Auxiliary data}

To provide constraints on the larger scale mass reservoir of detected disks, we make use of the ALMA 1.3~mm continuum map presented in \citet{Cheng18}. These data probe structures with sizes up to 18\farcs{6} (0.23~pc) with a spatial resolution of 1\farcs{07}$\times$1\farcs{02} (2600 AU). The 1~$\sigma$ noise level in this image is 0.45~\mjypbm. This low resolution millimeter map mainly traces the thermal dust emission from dense cores in the cloud. We refer to this map as the ``1\arcsec{}-resolution'' map, i.e., distinguishing it from the ``high-resolution'' map made with the long-baseline observation.

\section{Results} \label{sec:res}

\begin{figure}[ht!]
\epsscale{0.8}\plotone{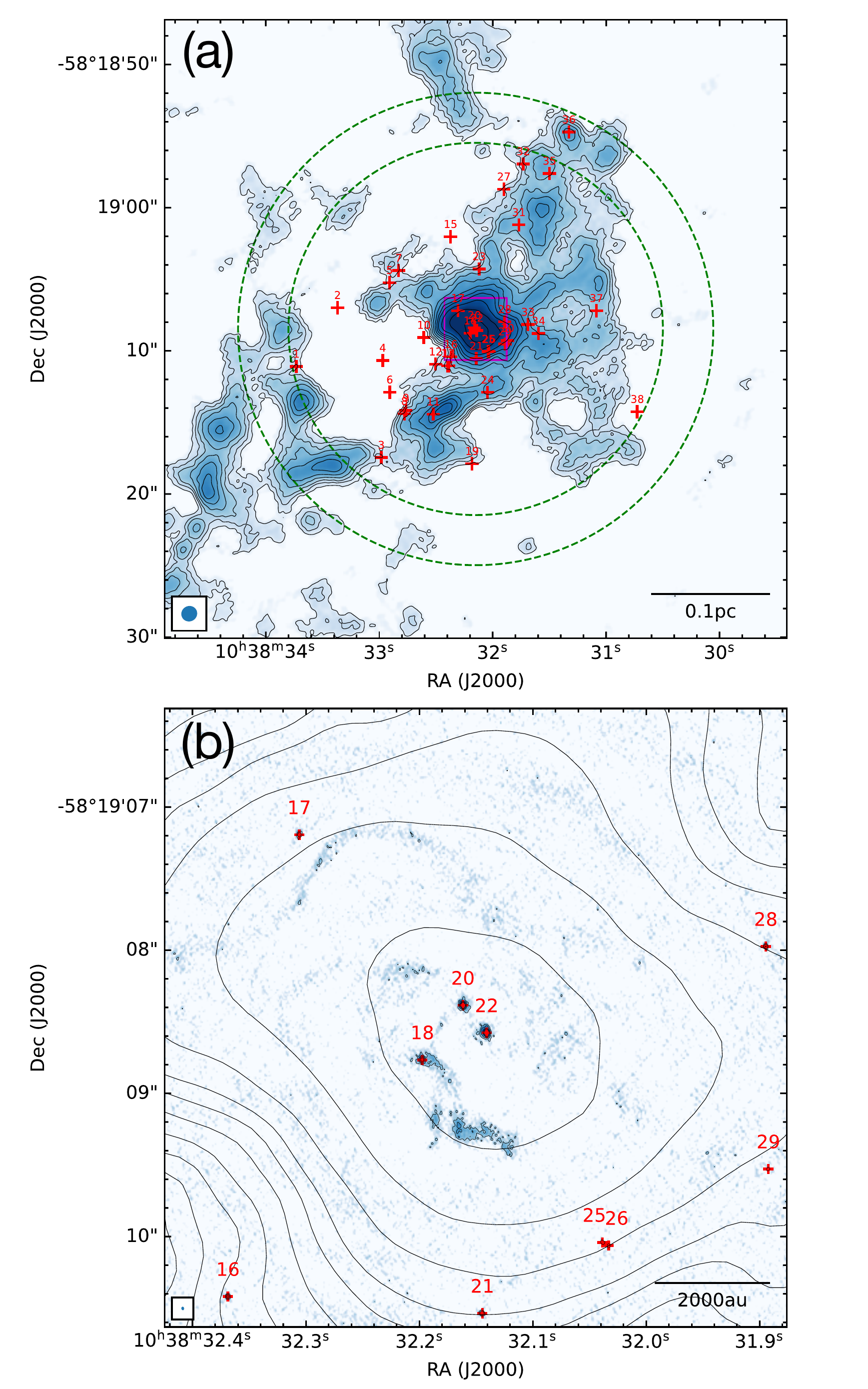}
\caption{{\it (a) Top:} Overview of compact continuum source detections in G286. Their positions are indicated with red crosses. The background is the low-resolution ($\sim$1\arcsec, or 2500~au) 1.3~mm continuum image from \citet{Cheng18}, shown in blue colorscale and contours. The contour levels are (4, 6, 8, 11, 15, 20, 25, 30, 40, 60, 100) $\times$ 4.5~\mjypbm. The beam size of the low resolution data, 1\farcs{12}$\times$1\farcs{07}, is indicated in the bottom left corner. The long-baseline observation is a single pointing centered on J2000 R.A.=10:38:32.2, Decl.=$-$58:19:8.5, and the inner and outer dashed green circles mark the boundaries where the primary beam correction factor corresponds to 0.5 and 0.3, respectively. The magenta box indicates the region shown in {\it (b)}. {\it (b) Bottom:} A zoom-in view showing the detections in the central 4\farcs{3} field of view. The blue colorscale and contours illustrate the long-baseline continuum image. The contour levels are (4, 8, 16, 32)$\times $ 15~\mujypbm. The black contours are the same as in {\it (a)}. The beam size of the long-baseline image, 26~mas$\times$20~mas, is indicated in the bottom left corner.}
\label{fig:overview}
\end{figure} 

In \autoref{fig:overview}a we present the high-resolution 1.3~mm continuum detections in the G286 region. The background is the 1\arcsec{}-resolution continuum image at the same wavelength, which illustrates the cloud fragments, i.e., dense cores, at $\sim$0.01~pc scales. The spatial resolution of long-baseline data is a factor of $\sim$50 improved (or a factor of 2500 improved in beam area), so the new observations probe physical structures at dramatically different scales, i.e., from $\sim$50 to $\sim$540~au. As we will show later, the high-resolution continuum is mainly tracing emission from protostellar disks, and the vast majority of detections are very compact or partly resolved point-like sources. The positions of these detections are marked in red crosses in \autoref{fig:overview}a. They are distributed mainly to the north and southeast of the central position. We enumerate the sources in order of increasing right ascension. 

In \autoref{fig:overview}b we show the high-resolution image in the central 4\farcs{5} field of view (FOV), i.e., the inner region of the most luminous dense core, G286c1, following the designation in \citet{Cheng20a}. The image reveals a triple system (Sources~18, 20, 22) in the center, as well as several other compact sources scattered around, with a distance about 1.5\arcsec{}--2.5\arcsec{} (3500~au -- 6200~au) from the center. Despite strong spatial filtering, some extended features are also seen. There are a couple of relatively diffuse condensations, located at $\sim$ 0\farcs{5} to the south of the triple system. They are detected at $>$ 4$\sigma$ level, but do not contain compact point sources. To the northeast of the triple system (separated by $\sim$1\farcs5) there is a tentative detection of a curved filamentary feature, which may be related to density enhancements shaped by complex gas motions on a few 0.01~pc scales. Its morphology is suggestive of an accretion stream to the central protostars.


\subsection{Source identification}
\label{sec:det}

We run the {\it dendrogram} algorithm \citep{Rosolowsky08} implemented in {\it astrodendro} to carry out an automated, systematic search for protostellar disks in the high-resolution continuum image. Here we focus on the identified leaf structures (the base element in the hierarchy of dendrogram that has no further sub-structure).
We tested different combinations of parameters and for the fiducial case we set the base flux density threshold to $5\sigma$, the minimum significance for structures to $1\sigma$, and the minimum area to 0.26$\times$ the synthesized beam. This set of parameters ensures the robustness of detections. The minimum area, i.e., 0.26~$\theta_{\rm beam}$, corresponds to the measured area in {\it dendrogram} for an isolated 6$\sigma$ point source (in {\it dendrogram} the area only accounts for pixels with values above the base level, i.e., 5$\sigma$ in our case). The search is run on the image before primary beam correction so that the rms noise is relatively uniform. However, the rms in the neighborhood of strong sources can be locally elevated due to imperfect cleaning and dynamic range limitations. In light of this, we first run {\it dendrogram} with a global rms $\sigma$ = 15~\mujypbm, then each detection is examined with the same criteria again with the rms replaced by a local value. The local rms is estimated in an annulus around each source from 0\farcs{1} to 0\farcs{2}, which ranges from 15 to 39~\mujypbm{} (after primary beam correction). Moreover, since we focus on the protostellar disks in this work, some relatively extended structures have been manually removed. This happens only in the central region shown in \autoref{fig:overview}b for the diffuse emission to the south of the triple system. 

\begin{figure}[ht!]
\epsscale{1.15}\plotone{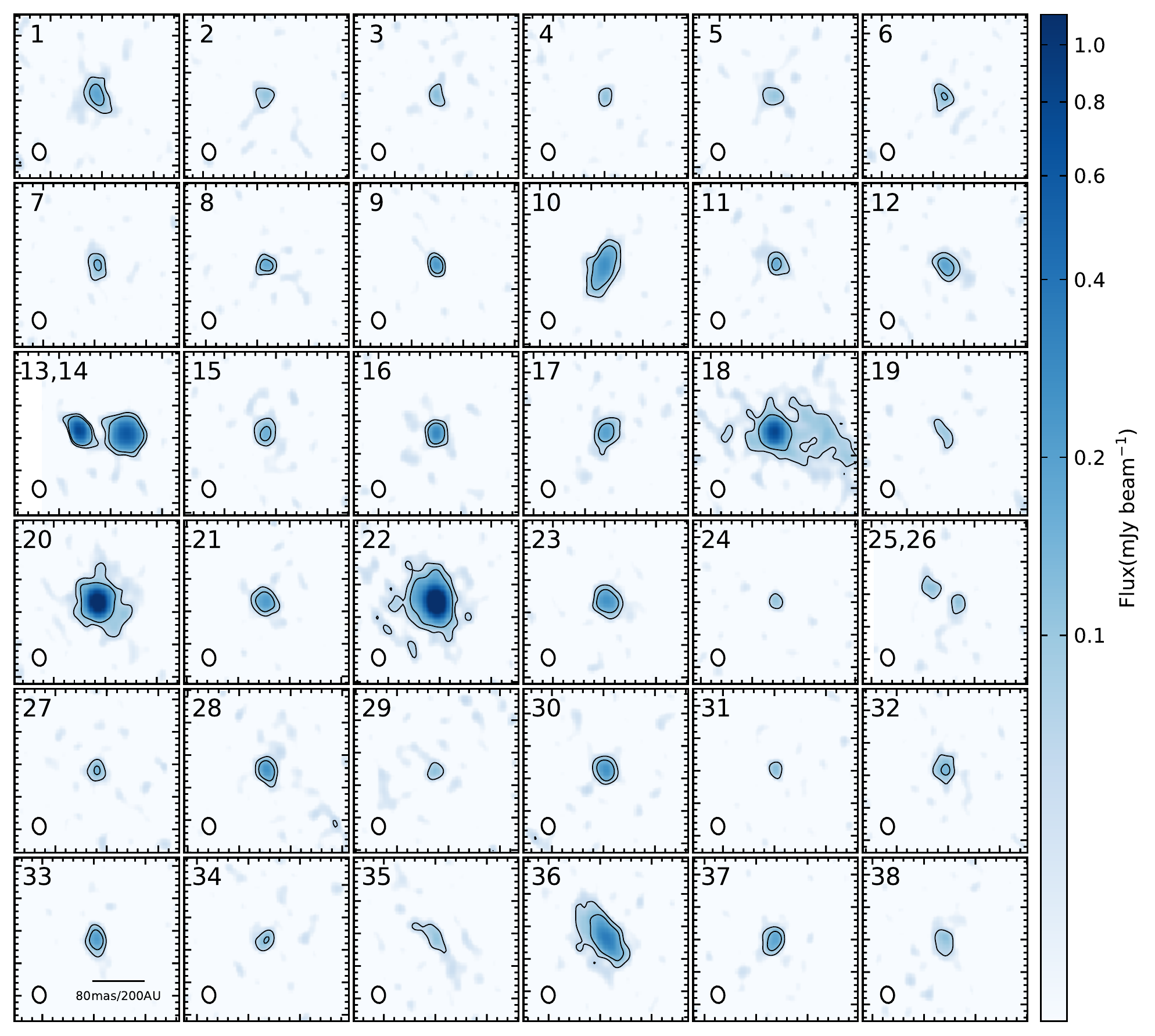}
\caption{{\it (a)} 1.3~mm continuum images of the protostellar disks in G286. The contours levels are (4, 8) $\times $ 15~\mujypbm. The beam size, 26~mas $\times$ 20~mas (65~au $\times$ 50~au), is shown in the bottom left corner of each panel. The close binary systems (sources 13 \& 14; sources 25 \& 26) are shown each within a single panel.}
\label{fig:disk}
\end{figure}

The identification gives a final catalog of 38 compact sources. In \autoref{fig:disk} we present close-up images for all the sources. In addition to the triple system in the center, we have also discovered three possible binary systems: sources~8, 9, separated by 620~au, sources~13, 14, separated by $\sim$150~au, and sources~25, 26, separated by $\sim$130~au. Since the region is highly clustered, we only consider multiple systems with projected separation $\lesssim$1000~au as candidate multiples. Most detections appear as unresolved or marginally resolved point-like sources, but some strong sources, including sources~10, 18, 20, 22, 36, are better resolved and some extended low-level emission is also seen. Sources~10 and 36 appear to have larger aspect ratios. We assume these detections are all associated with protostellar disks given their compact sizes ($\lesssim$200~au if located at 2.5~kpc). 

To examine the level of possible contaminants from extragalactic sources, we take the deep 1.2~mm ALMA survey of submillimeter galaxies in \citet{Gonzalez-Lopez20} as a reference \citep[see also][]{Fujimoto16,Aravena16,Arancibia22}. In this work the number density of submillimeter galaxies is 19300$^{+4700}_{-4400}$ per square degree for a flux density above 0.1~mJy. This would translate into a number of 0.79$^{+0.19}_{-0.18}$ in our FOV (inside FWHM of the primary beam) for $>$6$\sigma$($\sim$0.1~\mjypbm) detections. Thus we expect no more than one extragalactic contaminant in our sample.

\subsection{Characterization of the disks}\label{sec:cha}

To measure the fluxes and sizes of the disks from their 1.3~mm emission we fit 2D Gaussians using the {\it imfit} task in {CASA}. The results are listed in \autoref{table:disk}. Following \citet{Tobin20}, we adopt the 2$\sigma$ size of the deconvolved major axis as a proxy for the disk radius. If the intensity distribution is well described by a 2D Gaussian model, the 2$\sigma$ radius is approximately the radial point that contains 90\% of the total flux density in the curve-of-growth methodology employed by \citet{Ansdell16}. Six disks (sources 4, 8, 9, 13, 24, 31) are unresolved and could not have their deconvolved sizes constrained in the Gaussian fit. To derive an upper size limits for these sources, we run a series of experiments by generating synthetic disk images as a function of disk radius and signal to noise ratio (S/N), which are then used as input for Gaussian fit to determine the the minimum deconvolvable radius. We found that this radius is around 40~au for modest S/N of 7--30. The median radius in our sample, with upper limit measurements included, is 55~au. The first and third quartiles of the radius distribution are 43 and 67~au, respectively. There is a tendency for disks in multiple systems to be smaller, although the sample size is relatively small, i.e., 9 disks. These 9 disks all have disk radii $<$ 62~au, with a median $\lesssim$40~au. This trend is similar to that found in $\rho$~Oph, Taurus and Orion \citep[e.g.,][]{Cox17,Manara19,VanTerwisga22}, suggesting the formation (and evolution) of close multiple star systems affects the observable disk properties. 

We use the flux densities to calculate the mass of the protostellar disks, assuming that the emission purely comes from optically thin isothermal dust emission, enabling us to use the equation
\begin{equation}
    M_{\rm dust} = \frac{d^2F_\nu}{\kappa_\nu B_\nu (T_{\rm dust})}.
\end{equation}
where $d$ is the distance to the source, i.e., 2.5~kpc, $F_\nu$ is the observed flux density, $B_\nu$ is the Planck function, $T_{\rm dust}$ is the dust temperature and $\kappa_\nu$ is the dust opacity at the observed frequency $\nu$. We adopt $\kappa_{\rm 1.3 mm}$ = 0.899~$\rm cm^2g^{-1}$ from \citet{Ossenkopf94} (thin ice mantles, $\rm 10^6~cm^{-3}$ density). We multiply the calculated dust mass by 100, assuming a dust-to-gas mass ratio of 1:100 \citep{Bohlin78}, to obtain the gas mass. 
The average dust temperature we adopt for a protostellar system is assumed to be 20~K. If temperatures of 15~K or 30~K were to be adopted, then the mass estimates would differ by factors of 1.48 and 0.604, respectively. 
If a higher dust temperature of 100~K is assumed (suitable for typical massive sources \citep[e.g.][]{Zhang14}), then the mass estimate would decrease by a factor of 0.158.
Our measured dust disk masses range from 53 to 1825~$M_\oplus$, with a median of 172~$M_\oplus$. The first and third quartiles of the mass distribution are 136 and 261~$M_\oplus$, respectively. The corresponding total (gas) masses are from 0.016 to 0.55 $M_\odot$ with a median of 0.052 $M_\odot$. 

Note that different weighting schemes adopted in the imaging process could also affect the measured disk fluxes, especially for detections with modest statistical significance (S/N$\lesssim$10).
For example, in case of uniform weighting (20~mas$\times$14~mas resolution), bright disks are better resolved, but we fail to detect most disks, i.e., only 8 out of 38 disks still have peaks with S/N$>$6. The measured fluxes are systematically smaller, but consistent within 35\%. For the natural weighting image (31~mas$\times$25~mas resolution), the measured fluxes are larger than the fiducial values but mostly consistent within 25\%, with a few case larger by $\lesssim$50\%. This is partly due to the fact that the gaussian fitting is capturing more extended emission in lower resolution images, as evidenced by similarly larger deconvolved radii. 

\subsection{Correspondence with NIR and X-ray sources}\label{sec:cross_match}

To better characterize the properties of the disks, we check their correspondences with data at other wavelengths, which are listed in \autoref{table:disk}. First we cross-match our sample with NIR data. We use a catalog generated from the HST-WFC3/IR data presented in \citet{Cheng20b}. Source detections are done in the F110W and F160W bands down to a 4$\sigma$ limit, leading to a final catalog of $\sim$13,000 members inside a 6\arcmin$\times$6\arcmin{} field in G286, which also covers our FOV here (Cheng et al. in prep.). We found that 6 disks also exhibit a NIR counterpart within 0\farcs{05} (see \autoref{sec:nir}), i.e., sources 5, 13, 14, 24, 33, 35, suggesting that their association with relatively evolved YSOs or a viewing angle and local extinction conditions that enable escape of NIR light from the protostar. Here 0\farcs{05} roughly corresponds to the separations of close binary systems in the disk sample, and our HST catalog has position accuracy better than 0\farcs{05} even for relatively fainter sources.

We also search for infrared counterparts via a SIMBAD and Vizier coordinate search with a 1\arcsec\ radius. G286d8 and G286d9 appear to coincide with an infrared source 2MASS J10383269-5819143 \citep{Cutri03}. And the triple system G286d18/20/22 is associated with an infrared source G286.2086+00.1694 \citep{Mottram07} (see also the multi wavelength images in \autoref{sec:G286_sed}). These two infrared sources are resolved into some extended emission and/or are not centered on the disks in the F110W/F160W images (see \autoref{sec:nir}), and hence are not contained in our HST catalog. Given the limited angular resolution in relevant infrared surveys we are unable to explicitly identify the specific disk that corresponds to the infrared source.

We next cross-match our sample with an X-ray source catalog (from a census made with Chandra/ACIS observation (PI: Tan), Wang et al. in prep.). Pre-main-sequence (PMS) stars, both with and without disks, are well known to emit X-rays that can penetrate the heavy extinction of molecular cloud \citep{Feigelson07}, thus X-ray observations, particularly with the sub-arcsecond resolution with Chandra/ACIS, are highly efficient in selecting PMS stars in stellar clusters. We found 4 disks (source 2, 22, 24, 27) that coincide with X-ray sources within 0\farcs{5}, with one of them (source 24) also having a NIR counterpart. 

\subsection{Correspondence with dense cores}\label{sec:cross_match}

We also check the correspondence of disks with the dense cores identified in the 1\arcsec{}-resolution 1.3~mm continuum image, which traces the dust on 0.01~pc -- 0.2~pc scales, thus providing constraints on the surrounding environment of the disks. Class 0/I disks are expected to still reside within dense cores, while more evolved sources are not. 
Here we define that a disk is associated with a dense core if it is located within the core boundary defined by {\it Dendrogram} in \citet{Cheng18}. This criterion selects 16 disks. 

However, 11 of these are associated with G286c1, the massive ``core'' in the center of the FOV. Among these disks the triple system of sources 18, 20 and 22 appears to contain at least one actively accreting protostar: it is located near the emission peak of G286c1 (see \autoref{fig:overview}); shows some evidence for surrounding diffuse emission features that may be accretion streamers; and appears to be the source of a wide-angle bipolar outflow (Cheng et al. in prep.). The other 8 disks (source 17, 21, 25, 26, 28, 29, 30, 33) are also likely to still be embedded in dense, dusty gas, but it is unclear if they are all still actively accreting from this material. An intermediate resolution observation is needed to search for local concentration of mm emission around these sources. Only 5 disks that are associated with cores other than G286c1, i.e., source 7, 10, 18, 35, 36.

However, we notice that even for disks that still reside in dense cores, they appear to deviate from the emission peaks of cores, in contrast with the expectation for Class 0 disks. Some of these associations may be coincidental, i.e., due to projection effects.
If we use a more stringent criterion for cross matching between disks and cores, i.e., requiring disks to be located within 0\farcs{5} (1250~au) of emission peaks of dense cores, this will result in only two cases (source 36 and source 18, 20, 22).

{Note that the millimeter fluxes of Class 0 systems are envelope-dominated and should be strong enough to be detected in our 1\arcsec{} resolution image. To elaborate this we compare the G286 results with an ACA 0.87~mm survey of 300 protostars in Orion \citep{Federman22}, where it is found that over 80\% Class 0 systems have envelope fluxes greater than 200~mJy. This corresponds to a mass limit of 1.1~\msun, and translates into $\sim$2~mJy at 1.3~mm at a distance of 2.5~kpc, assuming a temperature of 20~K and a dust mass opacity law of $\kappa$ = ($\nu$/100 GHz) (\gcm). This is  marginally larger than the 4$\sigma$ level of the 1\arcsec{} resolution continuum ($\sigma$ = 0.45~mJy). Thus if we assume the G286 disk population has similar envelope properties as those in Orion, then in the low resolution map we should be able to detect the associated local peaks from envelopes for most protostars in the Class 0 stage. 
On the other hand, over 80\% Class I or Flat-spectrum systems have envelope fluxes smaller than 200~mJy, or masses lower than 1.1~\msun \citep{Federman22}, thus our 1\arcsec{} resolution observation is not sensitive enough to capture envelope fluxes from most protostars later than Class 0. }

Overall, the results suggest that most of our detected disks are Class I or later stage sources, with only a minority being Class 0. Still for a substantial fraction that do overlap with mm continuum emission in the 1\arcsec-resolution map, it is difficult to be certain of their evolutionary stage. Intermediate resolution mm continuum observations, as well as observations sensitive to the presence of outflows and dense gas tracers, will help resolve their status.

On the other hand,  the detections of circumstellar disks also inform us of the statuses of dense cores, i.e., if they are starless or protostellar. A related question concerns the absence of detected disks near the centers of many of the dense cores in G286 (see \autoref{fig:overview}). This could be due to these cores being pre-stellar in nature or having disks that are too faint (low-mass and/or low-temperature) to be detected by the ALMA long-baseline observation. Our 6$\sigma$ detection sensitivity corresponds to a gas mass of 0.012~\msun{} (or a dust mass of 41~$M_\oplus$) assuming 20~K temperature. Using the Orion protostellar disk survey in \citet{Tobin20} as a template, with such sensitivity we can detect $\sim$50\% disks in their sample (see \autoref{sec:comp}). Therefore, it is likely that some cores in G286 are forming stars with disks undetectable in this survey.




%

\begin{sidewaystable}

\startlongtable
\begin{deluxetable}{cccccccccccccc}
\tabletypesize{\tiny}
\tablecaption{Compact ALMA 1.3~mm continuum sources in G286\label{table:disk}}
\tablehead{
\colhead{Label} & \colhead{$\alpha$(J2000)} & \colhead{$\delta$(J2000)} & \colhead{$\rm FWHM_{decon}$} & \colhead{PA} & \colhead{rms} & \colhead{Peak} & \colhead{Flux} & \colhead{Dust mass} & \colhead{Gas mass} & \colhead{Radius} & \colhead{Dense Cores} & \colhead{HST F110W/F160W} & X-ray \\
\colhead{} & \colhead{(hh:mm:ss)} & \colhead{(dd:mm:ss)} & \colhead{(mas$\times$mas)} & \colhead{($^{\circ}$)} & \colhead{(\mujypbm)} & \colhead{(\mujypbm)} & \colhead{($\mu$Jy)} & \colhead{($M_\oplus$)} & \colhead{($M_\odot$)} & \colhead{(au)} &  &  & 
}
\startdata
G286d1  &  10:38:33.7310  &  $-$58:19:11.093  &  41$\times$25  &  26  &  32  &  361  &  977  &  443 & 0.133 & 87 & - & - & -   \\
G286d2  &  10:38:33.3676  &  $-$58:19:6.987  &  27$\times$19  &  $-$22  &  20  &  155  &  299  &  135 & 0.041 & 57 & - & - & Y   \\
G286d3  &  10:38:32.9820  &  $-$58:19:17.454  &  24$\times$13  &  12  &  25  &  196  &  306  &  139 & 0.042 & 51 & - & - & -   \\
G286d4  &  10:38:32.9697  &  $-$58:19:10.671  &  -  &  -  &  18  &  128  &  166  &  75 & 0.023 & $<$40 & - & - & -   \\
G286d5  &  10:38:32.9091  &  $-$58:19:5.241  &  39$\times$24  &  56  &  17  &  126  &  322  &  146 & 0.044 & 83 & - & Y & -   \\
G286d6  &  10:38:32.9077  &  $-$58:19:12.897  &  27$\times$18  &  14  &  17  &  161  &  301  &  136 & 0.041 & 57 & - & - & -   \\
G286d7  &  10:38:32.8296  &  $-$58:19:4.389  &  32$\times$15  &  7  &  18  &  171  &  319  &  145 & 0.043 & 68 & Y & - & -   \\
G286d8  &  10:38:32.7782  &  $-$58:19:14.394  &  -  &  -  &  19  &  214  &  286  &  130 & 0.039 & $<$40 & - & - & -  \\
G286d9  &  10:38:32.7652  &  $-$58:19:14.163  &  -  &  -  &  18  &  288  &  290  &  131 & 0.039 & $<$40 & - & - & -   \\
G286d10  &  10:38:32.6072  &  $-$58:19:9.078  &  57$\times$22  &  $-$24  &  16  &  279  &  1012  &  459 & 0.138 & 121 & - & - & -   \\
G286d11  &  10:38:32.5234  &  $-$58:19:14.424  &  34$\times$17  &  56  &  17  &  189  &  362  &  164 & 0.049 & 72 & Y & - & -   \\
G286d12  &  10:38:32.5036  &  $-$58:19:10.932  &  31$\times$19  &  52  &  15  &  194  &  411  &  186 & 0.056 & 66 & - & - & -   \\
G286d13  &  10:38:32.3958  &  $-$58:19:11.040  &  -  &  -  &  21  &  811  &  937  &  425 & 0.127 & $<$40 & - & Y & -   \\
G286d14  &  10:38:32.3871  &  $-$58:19:11.043  &  29$\times$27  &  $-$88  &  17  &  631  &  1593  &  722 & 0.217 & 62 & - & Y & -   \\
G286d15  &  10:38:32.3718  &  $-$58:19:2.028  &  29$\times$22  &  8  &  18  &  181  &  393  &  178 & 0.053 & 62 & - & - & -   \\
G286d16  &  10:38:32.3688  &  $-$58:19:10.419  &  15$\times$13  &  $-$46  &  16  &  310  &  409  &  185 & 0.056 & 32 & - & - & -   \\
G286d17  &  10:38:32.3059  &  $-$58:19:7.194  &  35$\times$20  &  $-$34  &  17  &  198  &  445  &  202 & 0.061 & 74 & Y & - & -   \\
G286d18  &  10:38:32.1974  &  $-$58:19:8.766  &  22$\times$21  &  78  &  24  &  795  &  1369  &  620 & 0.186 & 47 & Y & - & - \\
G286d19  &  10:38:32.1826  &  $-$58:19:17.883  &  48$\times$6  &  31  &  22  &  138  &  311  &  141 & 0.042 & 102 & - & - & -   \\
G286d20  &  10:38:32.1616  &  $-$58:19:8.385  &  14$\times$14  &  56  &  19  &  1943  &  2596  &  1176 & 0.353 & 30 & Y & - & - \\
G286d21  &  10:38:32.1445  &  $-$58:19:10.536  &  26$\times$19  &  61  &  15  &  219  &  432  &  196 & 0.059 & 55 & Y & - & -   \\
G286d22  &  10:38:32.1407  &  $-$58:19:8.577  &  16$\times$12  &  13  &  18  &  3051  &  4026  &  1825 & 0.548 & 34 & Y & - & Y \\
G286d23  &  10:38:32.1197  &  $-$58:19:4.284  &  26$\times$22  &  51  &  16  &  267  &  577  &  261 & 0.078 & 55 & - & - & -   \\
G286d24  &  10:38:32.0455  &  $-$58:19:12.900  &  -  &  -  &  17  &  110  &  117  &  53 & 0.016 & $<$40 & Y & Y & Y   \\
G286d25  &  10:38:32.0386  &  $-$58:19:10.041  &  25$\times$15  &  44  &  15  &  110  &  202  &  91 & 0.027 & 53 & Y & - & -   \\
G286d26  &  10:38:32.0333  &  $-$58:19:10.062  &  29$\times$17  &  $-$10  &  15  &  109  &  189  &  86 & 0.026 & 62 & Y & - & -   \\
G286d27  &  10:38:31.9019  &  $-$58:18:58.710  &  20$\times$19  &  $-$50  &  24  &  205  &  330  &  150 & 0.045 & 42 & - & - & Y   \\
G286d28  &  10:38:31.8947  &  $-$58:19:7.974  &  20$\times$13  &  27  &  19  &  257  &  370  &  168 & 0.050 & 42 & Y & - & -   \\
G286d29  &  10:38:31.8924  &  $-$58:19:9.528  &  22$\times$15  &  56  &  17  &  111  &  171  &  77 & 0.023 & 47 & Y & - & -   \\
G286d30  &  10:38:31.8695  &  $-$58:19:9.288  &  20$\times$17  &  59  &  18  &  260  &  417  &  189 & 0.057 & 42 & Y & - & -   \\
G286d31  &  10:38:31.7698  &  $-$58:19:1.200  &  -  &  -  &  20  &  151  &  136  &  62 & 0.019 & $<$40 & - & - & -   \\
G286d32  &  10:38:31.7306  &  $-$58:18:56.943  &  30$\times$21  &  1  &  27  &  254  &  564  &  255 & 0.077 & 64 & - & - & -   \\
G286d33  &  10:38:31.6894  &  $-$58:19:8.163  &  25$\times$11  &  4  &  15  &  225  &  347  &  157 & 0.047 & 53 & Y & Y & -   \\
G286d34  &  10:38:31.5965  &  $-$58:19:8.784  &  25$\times$14  &  $-$56  &  16  &  138  &  231  &  105 & 0.031 & 53 & - & - & -   \\
G286d35  &  10:38:31.4998  &  $-$58:18:57.609  &  60$\times$12  &  39  &  27  &  195  &  569  &  258 & 0.077 & 127 & Y & Y & -   \\
G286d36  &  10:38:31.3289  &  $-$58:18:54.720  &  52$\times$25  &  37  &  39  &  983  &  3432  &  1555 & 0.467 & 110 & Y & - & -   \\
G286d37  &  10:38:31.0862  &  $-$58:19:7.200  &  28$\times$16  &  $-$27  &  20  &  246  &  477  &  216 & 0.065 & 59 & - & - & -   \\
G286d38  &  10:38:30.7259  &  $-$58:19:14.249  &  41$\times$23  &  10  &  30  &  222  &  583  &  264 & 0.079 & 87 & - & - & -   \\
\enddata
\end{deluxetable}

\end{sidewaystable}


\section{Analysis and Discussion} \label{sec:dis}

\subsection{Comparison with the disk population in nearby regions}

\begin{deluxetable*}{lccccc}[!t]
\tabletypesize{\scriptsize}
\tablecaption{Disk Mass and Radius Sample Comparison \label{table:sample_comp}}
\tablewidth{0pt}
\tablehead{
\colhead{\multirow{2}{*}{Sample}} & \multicolumn{2}{c}{Mass} & & \multicolumn{2}{c}{Radius}  \\
\cline{2-3} \cline{5-6} &  \colhead{Logrank} &  \colhead{KS} &\colhead{}& \colhead{Logrank} & \colhead{KS} 
}
\startdata
G286 v.s. Orion (all) & 0.088 & 0.075 &  &0.049 & 0.17 \\
G286 v.s. Orion Class 0 & 0.0010 & 0.0022 &  &0.015 & 0.060 \\
G286 v.s. Orion Class I & 0.42 & 0.20 &  &0.25 & 0.50 \\
G286 v.s. Orion Class flat-spectrum & 0.58 & 0.16 &  &0.33 & 0.71 \\ \hline
G286 v.s. Taurus Class II & 0.013 & 0.022 &  &- & -\\
G286 v.s. Orion A Class II & 0.0072 & 0.012 &  &- & -\\
\enddata
\end{deluxetable*}

\begin{figure}[ht!]
\epsscale{1.1}\plotone{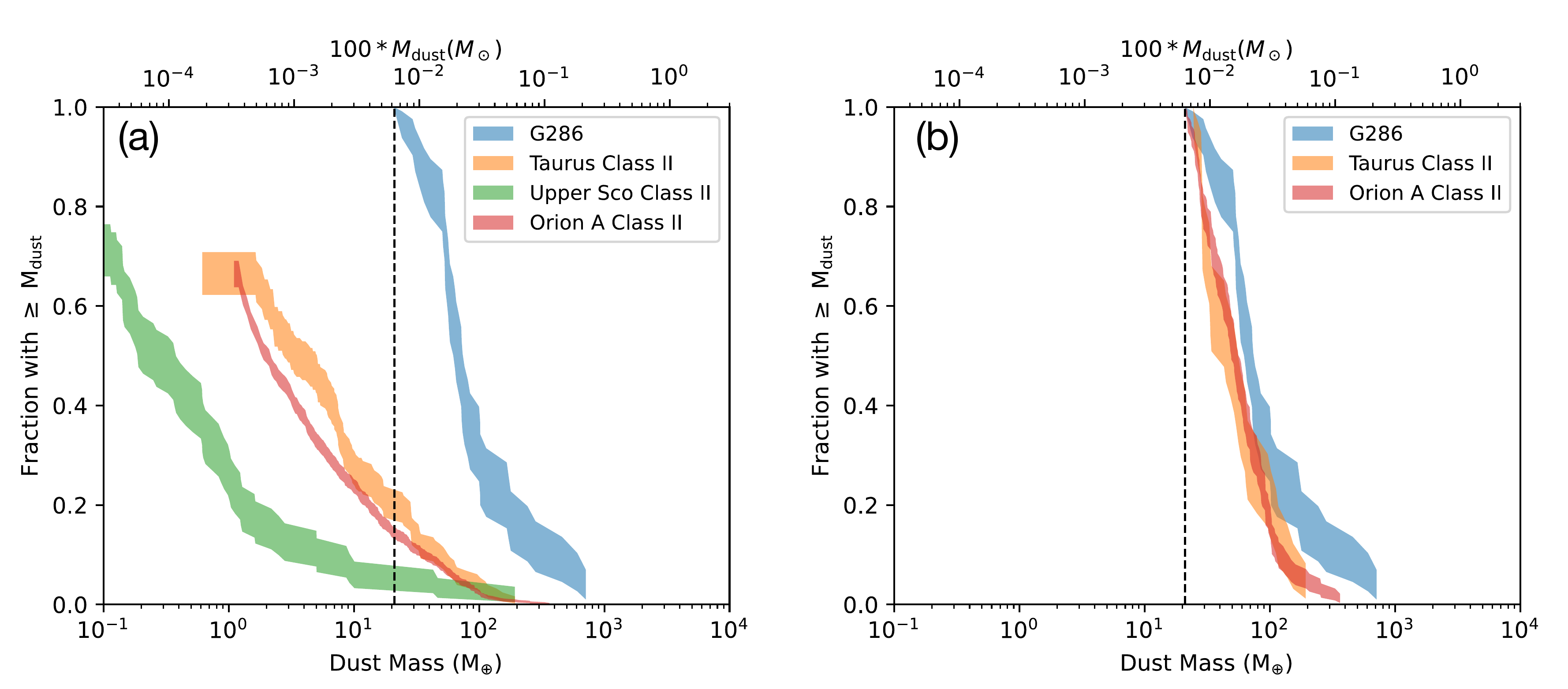}
\caption{{\it (a)} Cumulative distributions of dust disk masses in G286 compared to the Class II disk population in Taurus \citep{Tripathi17}, Upper~Sco \citep{Barenfeld16}, and Orion A cloud \citep{VanTerwisga22}. For comparison we recompute the disk masses in G286 using a uniform temperature of 20~K and a dust mass opacity law of $\kappa$ = ($\nu$/100 GHz)$^{\beta}$ (\gcm), where $\beta$ = 1. The dashed black line indicates the minimum detectable mass of 21~$M_\oplus$ in G286. The plots were constructed using survival analysis and the Kaplan-Meier estimator as implemented in the Python package {\it lifelines} \citep{Davidson-Pilon19}. The width of the cumulative distributions plotted represents the 1$\sigma$ uncertainty of the distribution. {\it (b)} Same as {\it (a)} but for Taurus and Orion A region we only include disks with dust masses exceeding the minimum detectable mass in G286, i.e., 21~$M_\oplus$. The distribution for Upper~Sco is not shown as there are only three disks that satisfy this criterion. 
}
\label{fig:comp}
\end{figure} 

\begin{figure}[ht!]
\epsscale{1.2}\plotone{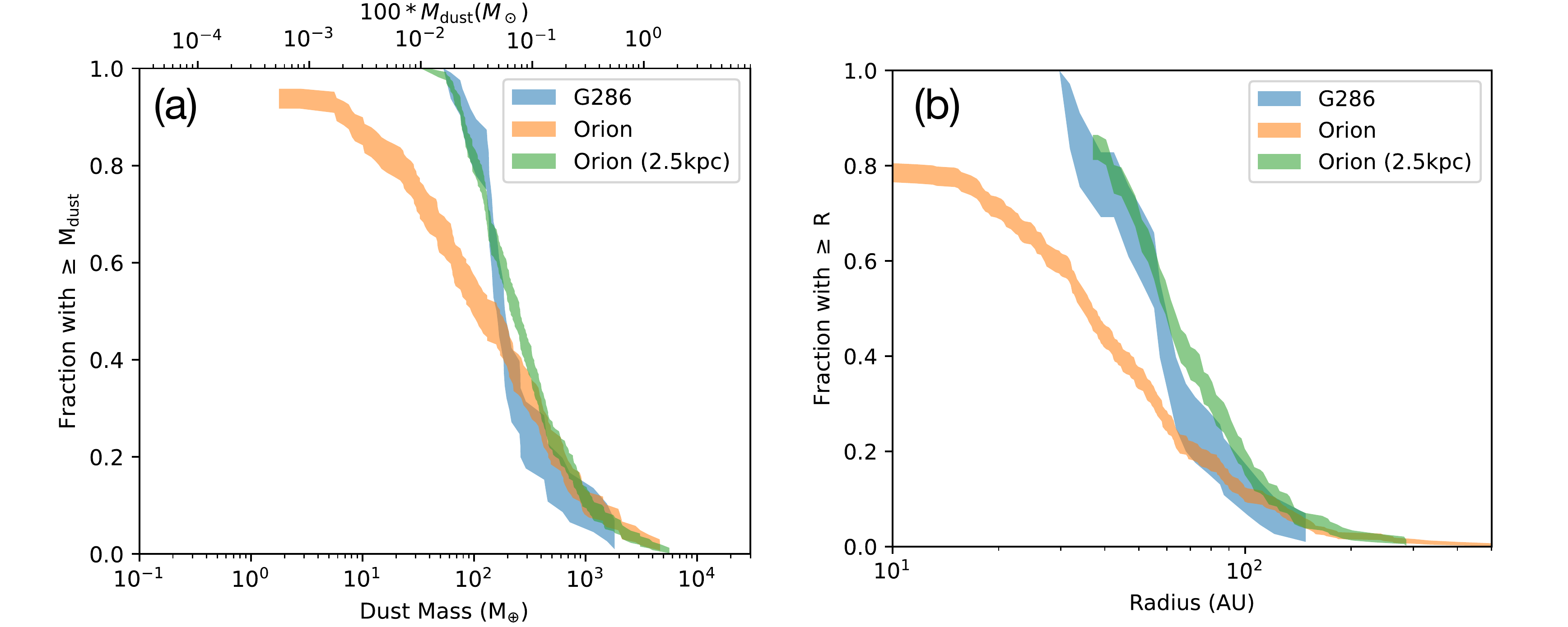}
\caption{{\it (a)} Cumulative distributions of dust disk masses in G286 compared with the Orion sample in \citet{Tobin20}. The mass distribution of the original Orion sample is shown in orange. The reprocessed Orion sample, after correcting for different target distance and observational setups, is shown in green. {\it (b)} Same as {\it (a)} but for disk radii. }
\label{fig:comp_orion1}
\end{figure} 

\begin{figure}[ht!]
\epsscale{1.2}\plotone{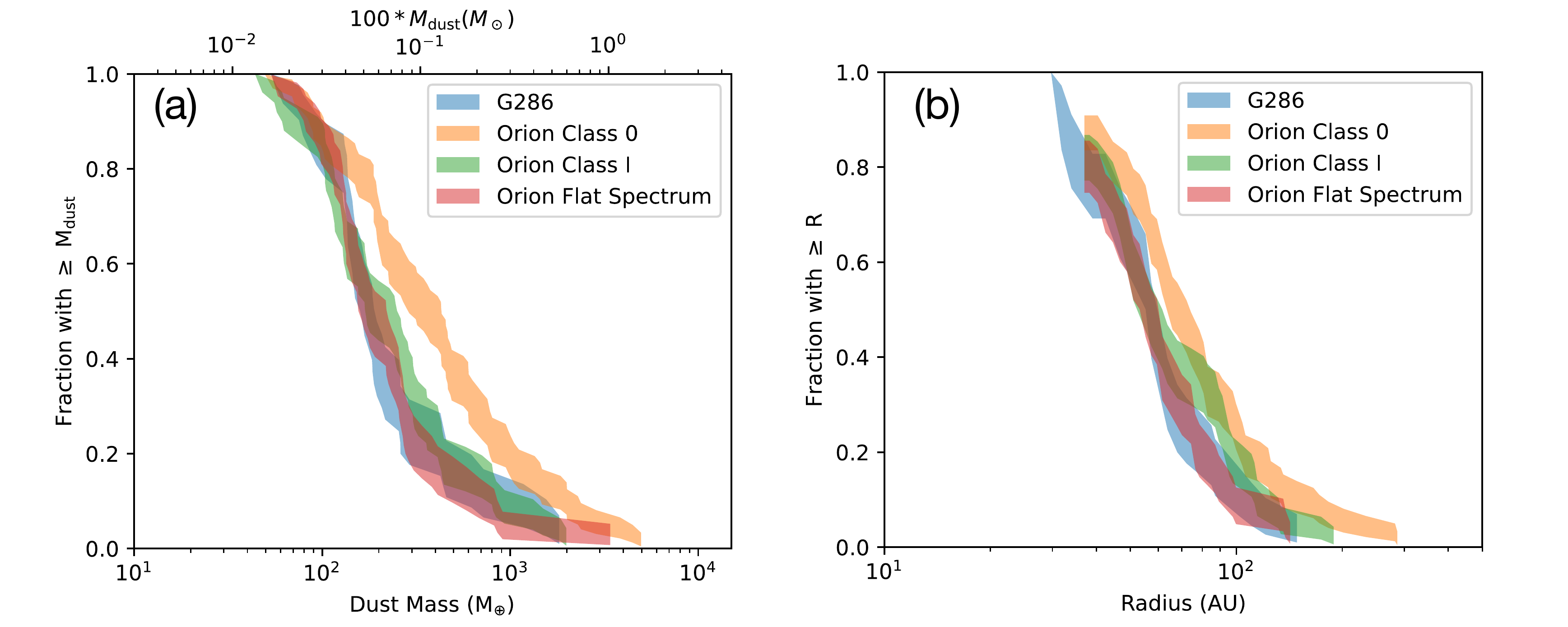}
\caption{{\it (a)} Cumulative distributions of dust disk masses in G286 compared with subgroups of simulated Orion sample in \citet{Tobin20} with different classifications. The Class~0 sources in Orion are significantly higher in mass with respect to G286. The Class~I and flat-spectrum protostars in Orion are in reasonable agreement with G286. {\it (b)} Same as {\it (a)} but for disk radii. Similarly the Class~0 sources in Orion are systematically larger compared to G286 while the Class~I and flat-spectrum protostars in Orion are similar to G286. }
\label{fig:comp_orion2}
\end{figure} 

\subsubsection{Comparison with Class II disks}

Our long-baseline observation provides a new opportunity to characterize the disk population in a massive protocluster at a distance of 2.5~kpc. The vast majority of the disks are newly detected and we do not have a prior knowledge of their classifications. We expect the sample contains mostly protostellar disks prior to Class II stage, since Class II disks are generally fainter and less likely to be detected with our sensitivity. To examine this, in \autoref{fig:comp} we compare the disk mass distribution of G286 with those of Class II disks in nearby regions. The disk sample in Taurus \citep{Tripathi17}, Orion~A \citep{VanTerwisga22} and Upper Sco \citep{Barenfeld16} are shown as representatives, with the former two being relatively young regions (1 -- 3~Myr), and the latter being an older region (5 -- 10~Myr). In general  Class II dust disks have systematically lower masses with increasing age of the stellar population, and hence the mass distribution of other regions with intermediate ages, such as Lupus and Chamaeleon~I, lie in between the selected regions \citep[e.g.,][]{Pascucci16,Ansdell16,Tobin20}. These Class II disk surveys usually adopt a uniform temperature of 20~K and a dust mass opacity law of $\kappa$ = ($\nu$/100 GHz)$^{\beta}$ (\gcm), where $\beta$ = 1. For comparison we recompute the disk masses in G286 with the same assumptions.

The dust disk mass distributions are shown as cumulative distributions using survival analysis implemented in the Python package {\it lifelines} \citep{Davidson-Pilon19}. The left censored fitting functions are used to account for upper limits derived from the nondetections, so the fraction does not reach unity at the low-mass end for Taurus, Orion~A, and Upper~Sco. For G286 the left censored data are not available since we do not have prior information on the expected positions of disks from other surveys, so the low mass part of the cumulative distribution is dominated by the sensitivity limit, but still the high-mass end ($M \gtrsim 100 M_\oplus$) should be relatively well constrained. We find that the disk mass distribution in G286 is systematically higher than those of Class II disks by a factor of $\gtrsim$3. 

To account for different sensitivity limits, we also show the cumulative distribution for subsamples of Taurus and Orion~A that only include disks with dust masses exceeding the minimum detectable mass in G286, i.e., 21~$M_\oplus$ (the distribution for Upper~Sco is not shown as there are only three disks that satisfy this criterion). The distinction is less dramatic when the observations are biased towards brighter sources, but the G286 disks are still systematically more massive. To establish the statistical significance of these differences, we use the two-sample log-rank test as implemented in {\it lifelines} and the results are presented in \autoref{table:sample_comp}. It is a nonparametric test for censored datasets to characterize the probability ($p$ value) that the two samples are randomly drawn from the same parent population. The mass distributions of G286 are inconsistent with being drawn from the same distributions as Taurus or Orion~A sub-sample ($p<$0.05). The probability values from the KS test are also shown as an additional check on the robustness of the log-rank test. 

Thus the G286 sample is less likely to be Class~II dominated, but should be mostly composed of protostellar disks. It is possible that the G286 sample contains a fraction of Class II disks. The minimum detectable disk mass in our sample is $\sim21\:M_\oplus$ (assuming 20~K and $\kappa_{\rm 1.3mm}$ = 2.3~\gcm). One can see that for Taurus, Orion~A, and Upper~Sco there exists a small fraction of Class II disks with masses exceeding this limit, i.e., $\sim$20\% of for Taurus and Orion~A, 5\% for Upper~Sco. Indeed, in \autoref{sec:cross_match} we have found four disks that are associated with X-ray emission, which are likely arising from Class II or Class I YSOs \citep[e.g.,][]{Feigelson99}. 

\subsubsection{Comparison with protostellar disks in Orion}\label{sec:comp}

We next compare the disk properties in G286 with those of the protostellar disk sample in the Orion molecular cloud presented in \citet{Tobin20}, which measures the disk properties with the same methodology and using similar wavelength data (0.87~mm) as our study. Again to account for different temperature assumptions in mass estimation, we recompute the disk masses in Orion using a uniform temperature of 20~K. As shown in \autoref{fig:comp_orion1}, the Orion disks have slightly larger masses than G286 at the high-mass end ($M \gtrsim 200~M_\oplus$). The two distributions significantly diverge at low-mass end due to different sensitivity limits. The radius distributions are broadly similar for disks with $r>60$~au, which can be better resolved in our observations, but in G286 we have not detected disks with radii greater than 150~au.


To account for the biases induced by different sensitivity and spatial resolution limits, we reprocess the Orion sample to simulate our observational setups in G286 with the following procedures. First, we take the Gaussian component (deconvolved FWHM$_{\rm major}$ and FWHM$_{\rm minor}$ axis, integrated flux density in 0.87~mm) of each disk in the Orion sample from \citet{Tobin20}. The major/minor disk axeses (in angular units) are then scaled by the distance ratio $d_{\rm Orion}$/$d_{\rm G286} = $0.17. The flux density is scaled from 0.87~mm to 1.3~mm according to the expectation of thermal dust emission,
\begin{equation}
    F_{\nu} \propto \frac{B_{\nu}(T_{\rm dust})\cdot \kappa_\nu}{d^2},
\end{equation}
where $F_\nu$ is the dust emission flux density at frequency $\nu$, $B_\nu(T_{\rm dust})$ is the Planck function with dust temperature $T_{\rm dust}$, $d$ is the target distance, $\kappa_\nu$ is the dust opacity, and we assume $\kappa_{\rm 0.87mm}$ = 1.84~\gcm{} and $\kappa_{\rm 1.3mm}$ = 0.899~\gcm{} \citep{Ossenkopf94}. The dust temperature has been estimated in \citet{Tobin20} based on a scaling relation with the bolometric luminosities. Then for each disk we can generate a simulated image by convolving the Gaussian component with our observational beam. Artificial Gaussian noise is also added (before convolution) to ensure that the final image has the same rms level as in our observations, i.e., 15~\mujypbm. Finally we run the same detection and characterization method described in \autoref{sec:det} and \autoref{sec:cha} on these simulated images to obtain the reprocessed Orion sample.

In this way we are able to detect $\sim$230 disks in Orion out of a sample of 477 disks, and their distributions in mass and radius are shown in \autoref{fig:comp_orion1}. The simulated Orion sample clearly has distributions that are in better agreement with those of G286 after correction for the observational biases. The Orion sample has a larger median mass of $235_{-117}^{+259}$~$M_\oplus$, compared to G286 ($172_{-36}^{+89}$~$M_\oplus$), but a similar median radius ($57_{-15}^{+26}$~au vs. $55_{-12}^{+12}$~au). Here the sub- and superscripts on the median values correspond to the first and third quartiles of the distributions. To establish the statistical significance of these differences, we use the two-sample log-rank test as implemented in {\it lifelines} and KS test, and the results are presented in \autoref{table:sample_comp}. We find that the mass and radius distributions of G286 are marginally consistent with being drawn from the same distributions as the simulated Orion sample at 90\% confidence ($p<$0.1). 
From the KS test the radius distributions between two samples are consistent with being drawn from the same distribution ($p$ = 0.17). 

We further compare the disks of G286 with those in subgroups of the simulated Orion sample with different classifications, i.e., Class 0, Class I and flat-spectrum. The cumulative distributions and statistical tests are presented in \autoref{fig:comp_orion2} and \autoref{table:sample_comp}, respectively. Interestingly, we find that for both masses and radii, the probability that the two distributions are drawn from the same distribution increases with the evolution stage of the subgroups in Orion, from Class~0 to Class~I/flat-spectrum. For the mass distribution the probability that two distributions are drawn from the same distribution is 1.0$\times$10$^{-3}$ for Class 0 group, 0.42 for Class I and 0.58 for flat-spectrum group, respectively. Similarly the radius distribution of G286 is inconsistent with that of Orion Class 0 sample ($p$ = 0.015), but statistically indistinguishable with the Class I and flat-spectrum sample ($p$ = 0.25, 0.33). 

A natural explanation is that the G286 sample has a low fraction of Class 0 disks and mainly consists of Class I/flat-spectrum disks. This agrees with the results in \autoref{sec:cross_match}, where we cross matched disks with dense cores. Most disks exhibit significant offsets relative to the continuum emission peaks on 0.01~pc scale, indicating they are not deeply embedded Class~0 objects. In this scenario there is no significant difference in disk properties between G286 and Orion, suggesting similar regulation in disk formation and evolution in the two regions. The fraction of Class 0 disks in G286 can be roughly estimated by 4/38$\sim$10\%, if we classify the four disks that are close to emission peaks of dense cores (i.e., sources 18, 20, 22, 36) as Class~0 stage (or analogous to Class~0 for high-/intermediate-mass YSOs). This is significantly lower than the fraction in Orion, i.e., 94/328 $\sim$29\% \citep{Tobin20}.

{The G286 protocluster is characterized as being a highly clustered environment, with a high gas mass surface density ($\gtrsim$ 0.3~\gcm{} for the central 15\arcsec{} FOV, \citeauthor{Cheng20a} \citeyear{Cheng20a}), making it different from most regions in Orion and other nearby star formation regions. In spite of different physical conditions, the differences in disk mass/radius distribution between G286 and Orion are potentially explained as being due to a variation in evolutionary status, i.e., relative fraction of Class 0/I/flat-spectrum objects and there is no indication for systematic difference in disk properties.

Similarly, \citet{Tobin20} examined subgroups of disks within Orion that have different environmental conditions, such as L1641 and the Integral-Shaped Filament region, and found no significant variations. Surveys in the Perseus also reveal a similar mass distribution of Class I disks as in Orion, although the Class 0 disks appear to be more massive \citep{Tychoniec20}. In contrast, the Orion protostars of all classes have systematically higher disk dust masses than those in the Ophiuchus \citep[e.g.,][]{Cieza19,Encalada21}, but this may be due to a sample classification issue instead of a true difference \citep[see][for a discussion]{Tobin20}. 

The similarity in disk dust properties among regions may suggest limited environmental dependence of disk properties. However, one may wonder if such consistency is universal, especially when some environmental factors that are known to be important in regulating disk formation and evolution such as strength of magnetic field \citep[e.g.,][]{Li14}, are taken into account. Further surveys covering more environments are needed to shed more light on this issue. Moreover, our analysis here also highlights the importance of uniform observational setups and methodologies when performing comparisons among different regions.
}
\subsection{Disk properties relative to the cluster center}

\begin{figure}[ht!]
\epsscale{1.1}\plotone{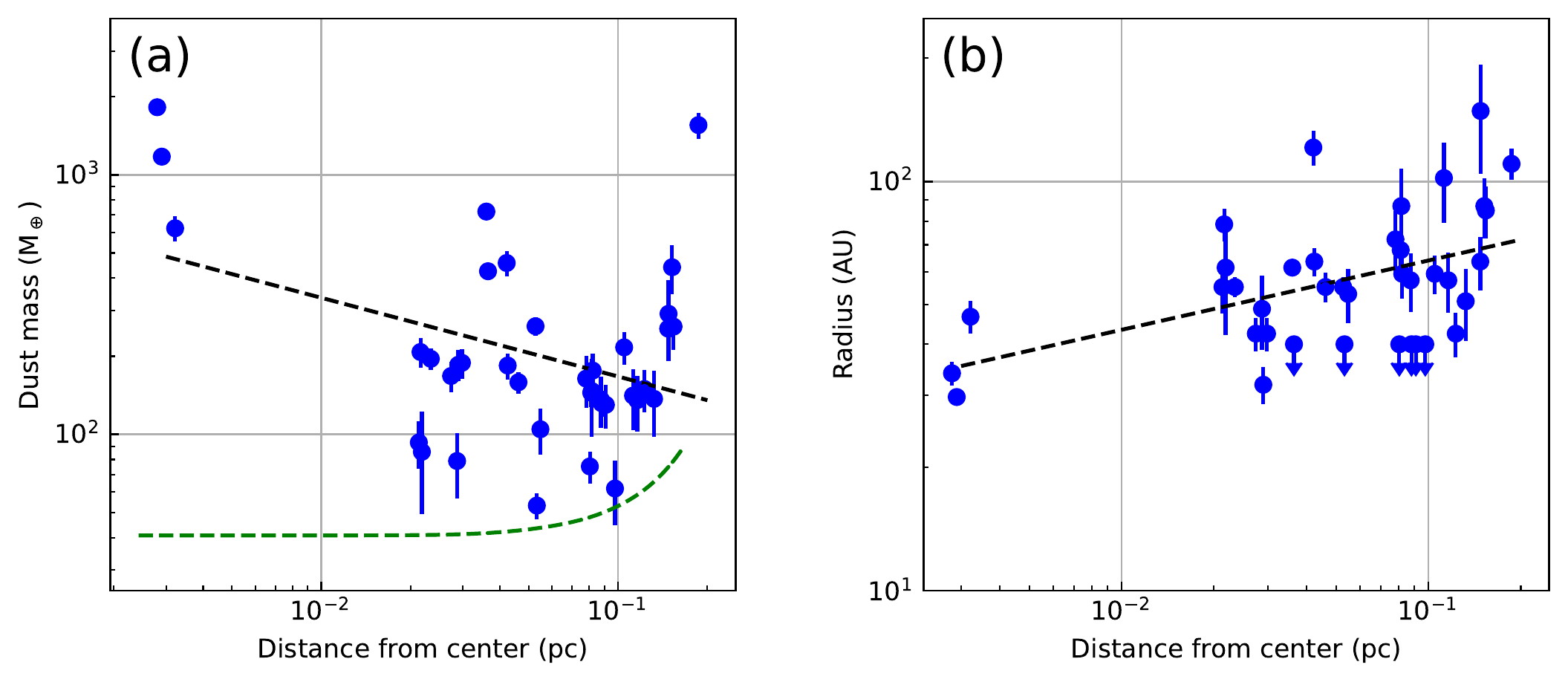}
\caption{{\it (a)} Scatter plot of disk mass against the distance from cluster center (see text for details). The black line shows a linear fit to the data. The green curve indicates the 6$\sigma$ detection threshold. The rms noise varies with distance from the cluster center, which is also the phasecenter of the observation, due to beam response. The errorbars only account for uncertainties from Gaussian fitting.} The typical flux uncertainty from Gaussian fitting is 10\% -- 25\%. The disk mass estimation also relies on assumptions of the temperature and dust properties. {\it (b)} Same as {\it (a)} but for disk radii. The typical uncertainties for the radius estimation are 10\% -- 20\% from the Gaussian fitting.
\label{fig:dist}
\end{figure} 

\begin{deluxetable*}{lcccccc}[!t]
\tabletypesize{\scriptsize}
\tablecaption{Statistical tests for correlations between the disk properties and the distance from cluster center \label{table:dist_tst}}
\tablewidth{0pt}
\tablehead{
\colhead{\multirow{2}{*}{Sample\tablenotemark{a}}} & \colhead{\multirow{2}{*}{Number of disks}} & \multicolumn{2}{c}{Mass v.s. Distance} & & \multicolumn{2}{c}{Radius v.s. Distance}  \\
\cline{3-4} \cline{6-7} &  & \colhead{Spearman $\rho$}  & \colhead{$p$-value} & \colhead{} &\colhead{Spearman $\rho$}& \colhead{$p$-value}  
}
\startdata
G286 (all) & 38 & $-$0.10 & 0.554 &  & 0.55 & 0.001 \\
G286 (Distance $>$ 0.01pc) & 35 & 0.13 & 0.454 &  &0.41 & 0.026 \\
G286 (Distance $<$ 0.15pc) & 35 & $-$0.30 & 0.081 &  &0.43 & 0.019 \\
G286 (nonmulitple) & 29 & 0.14 & 0.477 &  &0.45 & 0.023 \\
\enddata
\tablenotetext{a}{We consider four samples for the statistical tests: the total 38 disks; 35 disks with the massive triple system in the center removed (i.e., Distance $>$ 0.01~pc); 35 disks excluding the three detections where the beam response is smaller than 0.5 (i.e., Distance $<$ 0.15~pc); 29 disks that are not contained in multiple systems (see \autoref{sec:det}). }
\end{deluxetable*}

We check for possible trends between disk properties and the separation from the cluster center. Here the cluster center is defined as the position of dense core G286c1, i.e., the emission peak of the 1\arcsec{}-resolution 1.3~mm continuum map. This is a reasonable choice from a cluster formation perspective, since G286c1 is located at the central hub where two large scale filaments intersect \citep{Cheng20b}. Furthermore, it is the most massive dense core in G286 and hosts a massive multiple system in formation. We have also attempted to define the cluster center in other ways, e.g., using the average position of disks/dense cores, and found that it does not affect the main conclusion of this section. 

In \autoref{fig:dist} we show scatter plots of disk mass and disk radius versus projected distance from the cluster center. There is a lack of disks for distances from 0.004~pc to 0.02~pc. This is similar to the distribution observed in the GGD~27 cluster for which \citet{Busquet19} suggests may be caused by disks being impacted by a central massive star(s) or central crowded cluster environment.
From \autoref{fig:dist} there appear to be weak trends for lower disk masses and larger disk radii with increasing projected distance. A linear fit in log-log space gives the following relations:
\begin{equation}
\textrm{log} (M/M_\oplus) = (-0.30 \pm 0.12)~\textrm{log} (D/\textrm{pc}) + (1.92\pm0.17),
\end{equation}
and
\begin{equation}
\textrm{log} (R/\textrm{au}) = (0.17 \pm 0.05)~\textrm{log} (D/\textrm{pc}) + (1.98\pm0.07).
\end{equation}
While there is large scatter seen in the figures, the trends appear to have potential significance, i.e., with the indices being $\gtrsim 3\sigma$ from a flat relation.

To further check the strength of the correlations, we use the Spearman rank coefficient test, which assesses how well the relationship between two variables can be described using a monotonic function. The Spearman test returns a value $\rho$ between $-$1 and $+$1 and an associated $p$-value to assess the significance of the correlation. A Spearman correlation of 1 results when the two variables being compared are positively related, whereas negative values of $\rho$ represent a negative correlation. 

As summarized in \autoref{table:dist_tst}, there is no significant correlation between mass and distance, with $\rho$ = $-$0.10 and $p$=0.554. This negative correlation becomes stronger and marginally significant ($\rho$=$-$0.30, $p$ = 0.081) if we exclude the three disks with distance greater than 0.15~pc, where the detections are more severely affected by beam response. However, the triple system in the center is associated with more massive YSOs and thus presumably has disk temperatures higher than 20~K, leading to overestimated disk dust masses. If the triple system is excluded, we see no clear trend between mass and distance from \autoref{table:dist_tst} (and \autoref{fig:dist}).

On the other hand, there is a modest correlation between radius and distance ($\rho$ = 0.41--0.55), which seems to be relatively robust ($p$ = 0.001 -- 0.026 for different subsamples). This correlation remains for disks that are not contained in multiple systems ($\rho$ = 0.45, $p$ = 0.023). 

One environmental factor that could result in a radial dependence of disk properties is photoevaporation from massive stars, if these tend to be located at the cluster center. 
Incident EUV radiation fields are expected to reduce the masses and radii of protoplanetary disks by photoevaporating the gas \citep[e.g.,][]{Hollenbach00,Nicholson19}, leading to a decreasing trend in disk masses/sizes with proximity to the ionizing star. Such an effect has been claimed in the ONC \citep{Mann14,Eisner18}, however deeper disk surveys have disputed this result \citep{Otter21}. 
While the observed trend of radii in G286 is seemingly in line with the prediction of photoevaporation, we do not expect it to be an effective mechanism, because the ionizing flux is much weaker than in ONC. The stellar mass of G286c1 is not well constrained from its spectral energy distribution (SED) from mid to far infrared wavelengths (see \autoref{sec:G286_sed}). If we assume the fluxes are contributed by a dominating component in the triple system, a range of protostellar properties, i.e., a central mass of 4 -- 12~\msun, can yield SEDs that are consistent with the observation. But in any case, G286c1 is still in an early embedded phase and has not developed a large HII region. 




Disk truncation by dynamical interactions is another process that could lead to a radial dependence of disk mass and disk size with radial location in the cluster. The increasing trend of disk radii with distance could indicate that strong dynamical interactions in the central region are more common than in the cluster outskirts. Alternatively, the trend may reflect inherited properties of protostellar cores, e.g., if the initial cores tend to be smaller in the higher pressure central regions, i.e., as expected in the Turbulent Core model \citep{McKee03}, and if disk sizes are related to initial core sizes.

Finally, we note that there are relatively large uncertainties in the deconvolved sizes from Gaussian fits in relatively low S/N cases (S/N$<$10), and it is likely that we overestimate the dust disk size when the disk emission becomes confused by background noise, e.g., in case of source 35. A systematic variation of noise properties with radial position in the cluster could thus also influence these results. A more sensitive and complete survey of the G286 disk population is needed to further explore this tentative trend of disk radii growing with radial position in the cluster.

\subsection{Mass segregation}

\begin{figure}[ht!]
\epsscale{1.1}\plotone{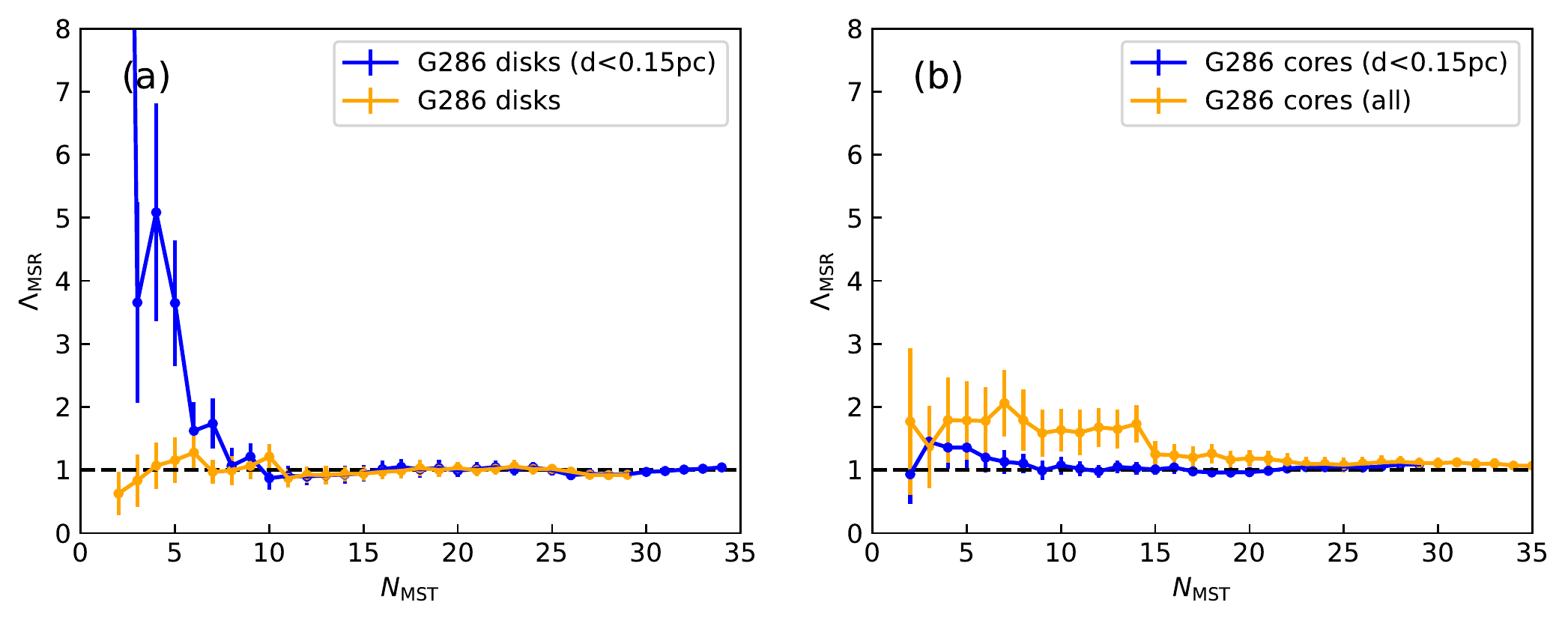}
\caption{{\it (a)} Mass segregation ratio $\Lambda_{\rm MSR}$ as a function of the $N_{\rm MST}$ most massive disks of G286. The blue line indicates the results when we use only the sources within 0.15~pc from the cluster center, while the the orange line indicates the results for all the disks listed in \autoref{table:disk}. {\it (b)} Mass segregation ratio $\Lambda_{\rm MSR}$ for the dense cores in G286. The blue line indicates the results when we use only the sources within 0.15~pc from the cluster center, while the the orange line indicates the results for all the disks cataloged in \citet{Cheng20b}.}
\label{fig:segre}
\end{figure} 

We further examine if there is signature for mass segregation in the disk population, i.e., if the most massive disks (inferred by mm continuum emission) are more centrally concentrated and/or clustered than expected from a random distribution. If there is a correlation between disk mass and protostar mass, then such a mass segregation signal could be due to stellar dynamical mass segregation. Such mass segregation is observed in dynamically old stellar clusters, but also in very young regions, suggesting the segregation is at least partially primordial \citep[e.g.,][]{Meylan00,Gennaro11}. Observations of 0.002~pc--0.1~pc scale dense cores, i.e., the progenitor of single stars or small multiple systems, seem to support this argument, since at least a modest level of segregation is found in many molecular clouds \citep{Kirk16,Plunkett18,Parker18,Dib19,Sadaghiani20,Nony21}. However, this is not true for all star forming regions \citep[e.g.,][]{Dib19,Sanhueza19}. It is still unclear when and how such primordial segregation develops and how its level evolves from dense cores to the YSO stage with the impact of fragmentation and further accretion \citep{Alcock19}. If we assume a correlation between the disk mass and stellar mass \citep[e.g.,][]{Pascucci16}, then the current mass segregation level of protostars in G286 can be estimated by examining the spatial distribution of the disk population.

Following \citet{Allison09} the mass segregation ratio ($\Lambda_{\rm MSR}$) is
\begin{equation}
    \Lambda_{\rm MSR} = \frac{L_{\rm norm}}{L_{\rm massive}} \pm \frac{\sigma_{\rm norm}}{L_{\rm massive}},
\end{equation}
where $L_{\rm massive}$ is the average path-length of the minimum spanning tree (MST) of the $N$ most massive sources, and $L_{\rm norm}$ is the average path-length of the MST of $N$ random sources in the cluster. We take the average over 1000 random sets of $N$ sources in the cluster to estimate $L_{\rm norm}$ and its statistical deviation. A $\Lambda_{\rm MSR}$ greater than unity indicates a concentration or clustering of massive sources with respect to the random sample. The larger the $\Lambda_{\rm MSR}$, the more mass segregated the sample. 

\autoref{fig:segre} displays (with orange lines) the mass segregation ratio as a function of number of mass-ranked members ($N_{\rm MST}$). The diagram shows values of $\Lambda_{\rm MSR}$ close to unity, suggesting no mass segregation in the sources as ranked by protostellar disk mass. This is partly due to the existence of source 36, which is located close to the edge the FOV, but has the second largest mass among the sample, thus driving $L_{\rm massive}$ to a very large value. If we only account for sources with d$<$0.15~pc, i.e., with source 36 removed, then we have a large $\Lambda_{\rm MSR}$ of 32 for $N_{\rm MST}$ = 2, and $\Lambda_{\rm MSR}\gtrsim$3 for $N_{\rm MST}\leq$ 5. Therefore, in the inner region, i.e., excluding source 36, there is evidence for mass segregation.
One caveat is that the dust disk masses for sources 18, 20 and 22 could be overestimated if the disk temperatures are much higher than the assumed 20~K given that they are associated with high-/intermediate mass star formation (\autoref{sec:G286_sed}).

We have also calculated the $\Lambda_{\rm MSR}$ parameter for the dense core sample in G286 using the catalog in \citet{Cheng20b}, as shown in \autoref{fig:segre}. We first limit the core sample to have a similar spatial range as the disks, i.e., within 0.15~pc from the cluster center. This gives a $\Lambda_{\rm MSR}$ around 1, suggesting no significant mass segregation. If all the cores in \citet{Cheng20b} are used, then there appears to be a weak trend of mass segregation, with $\Lambda_{\rm MSR}\sim$1.5 for $N_{\rm MST}<$ 15. Therefore, overall there is no obvious mass segregation of cores in G286.

In summary, for both dense core and disk population in G286, there is no strong evidence for widespread mass segregation, but high-mass star formation as evidenced by the example of G286c1 and its associated protostellar sources, is located at the cluster center.
Our disk sample is mainly composed of Class I/flat-spectrum objects that have been decoupled from their parental cores. Assuming the current observed dense core sample has similar properties as the cores that host these disks, the observation indicates that there is no significant variation in the mass segregation level from cores to protostars. 

We also note that the scaling relation between disk dust mass and stellar mass is mainly established for Class~II disks \citep[e.g.][]{Pascucci16,Ansdell16,Ansdell17}, which may originate from a mixture of both the initial conditions and the evolutionary process \citep{Manara22}. It is less clear if such correlation still hold for protostellar disks. Qualitatively disk masses do tend to be greater for more massive protostellar systems in the large-scale hydrodynamics disk population synthesis \citep{Bate18}. A tighter correlation is expected in simulations that render gravitationally self-regulated protostellar disks that are in a marginally gravitationally unstable state \citep{Xu21,Xu22}. However, from the observational side this correlation has been less studied. Considering the large dispersion in this relation even for Class~II disks \citep[$\sim$0.6--0.9~dex in disk dust mass values for a given stellar mass, see][]{Manara22}, the segregation in disk dust mass may not robustly reflect that of the protostar population. Nevertheless, the properties of this segregation are an observable that can be compared to simulations that aim to predict disk properties in protocluster environments.

Another limiting factor here is that our disk detections are limited to a relatively small FOV, i.e., within $\sim$0.15~pc from the center, and the rms noise is higher with larger distance from center due to beam response, thus hindering detection of low mass disks at the cluster outskirts. Again, a more complete and deeper disk survey would help to explore the question of mass segregation more fully.

\section{Conclusions}\label{sec:sum}

We have utilized the long-baseline capability of ALMA to conduct a survey of protostellar disks in the massive protocluster G286.21+0.16 at a distance of 2.5~kpc. With a resolution of 23 mas (58~au) and a sensitivity of 15~\mujypbm (0.002~\msun, assuming 20~K), we detected 38 compact continuum sources in the 1.3~mm continuum, most of which should be tracing protostellar disks. These disks have dust masses ranging from 53 to 1825 $M_\oplus$ assuming a temperature of 20~K. The median dust disk mass and radius are 172~$M_\oplus$ and 52~au, respectively. Among the sample there is a triple system located at the center of the cluster, as well as three binary systems. 

We have suggested that this sample is mainly composed of Class I/flat-spectrum disks, since they are typically not closely associated with dense cores as expected for disks in Class 0 stage, while Class II sources would generally be too faint to be detected.
Also, we have found that there is no statistical difference in the distributions of disk masses and radii between the G286 sample and the Class I/flat-spectrum objects in Orion molecular cloud, indicating that disk formation and evolution in these different regions undergoes similar regulation. 

We have found a tentative trend of increasing disk radius with projected distance from the cluster center. Photoevaporation is not expected to be the cause of such a trend, leaving dynamical interactions or inherited properties from natal cores as possible explanations.
We did not find strong evidence for mass segregation in both the disk and dense core population. However, if we restrict to the inner 0.15~pc, then there is some evidence for mass segregation in the disk population. In any case, the most massive core and its associated massive disk systems are located in the center of the cluster. The detailed properties of this massive protostellar system will be explored in more detail in a follow-up paper, including analysis of the line emission from this ALMA observation.


This work demonstrates the capability to characterize the disk populations in embedded protoclusters at distances as far as 2.5~kpc with ALMA at 1.3~mm. If shorter wavelength, e.g., 0.9~mm, are used, then similar spatial resolutions ($\sim$50~au) can be achieved for targets at $\sim$4~kpc. More distant regions could also be feasible if spatial resolution is not particularly important. This allows for characterization of protostellar disks, as well as multiplicity, in regions that span a wide range of physical conditions. The presented work can potentially serve as a template for surveys of larger samples to study the environmental dependence of disk properties.

\newpage
\appendix
\counterwithin{figure}{section}
\counterwithin{table}{section}

\section{Disk counterparts at NIR wavelength}\label{sec:nir}

\begin{figure*}[ht!]
\epsscale{0.9}\plotone{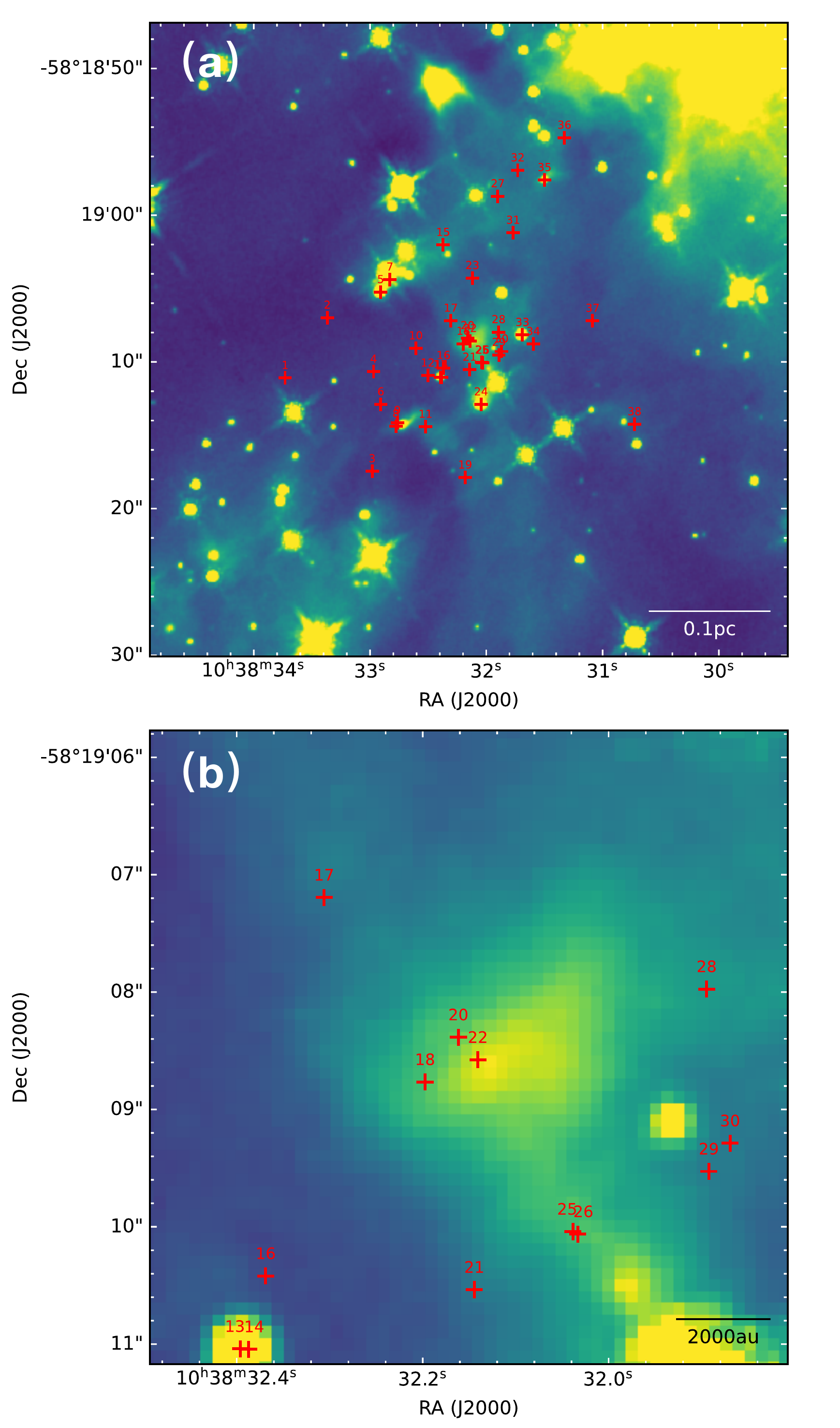}
\caption{ {\it (a)} F160W map of G286 observed with HST-WFC3/IR. The positions of disks are indicated with red crosses. {\it (b)} A zoom-in view of the central 5\farcs{4} field of view.} \label{fig:nir1}
\end{figure*}

\begin{figure*}[ht!]
\epsscale{1.0}\plotone{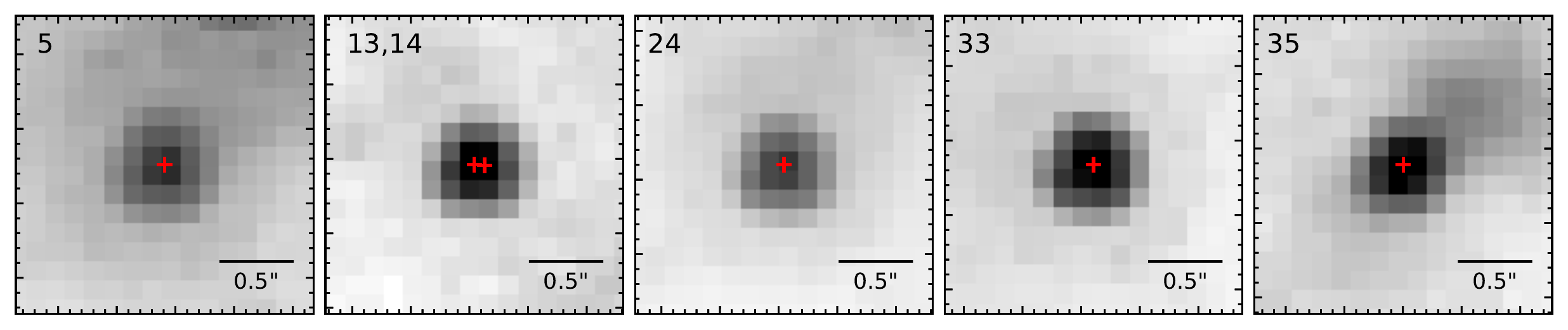}
\caption{HST F160W image of 6 disks with NIR counterparts in a zoom-in view. The indexes of sources are labeled at the top left corner.\label{fig:nir2}
}
\end{figure*}

In \autoref{fig:nir1} and \autoref{fig:nir2} we present the HST F160W map of G286, and zoom-in views for disks with NIR counterparts, respectively.

\section{SED fitting of G286c1}\label{sec:G286_sed}

\begin{figure*}[ht!]
\epsscale{1.1}\plotone{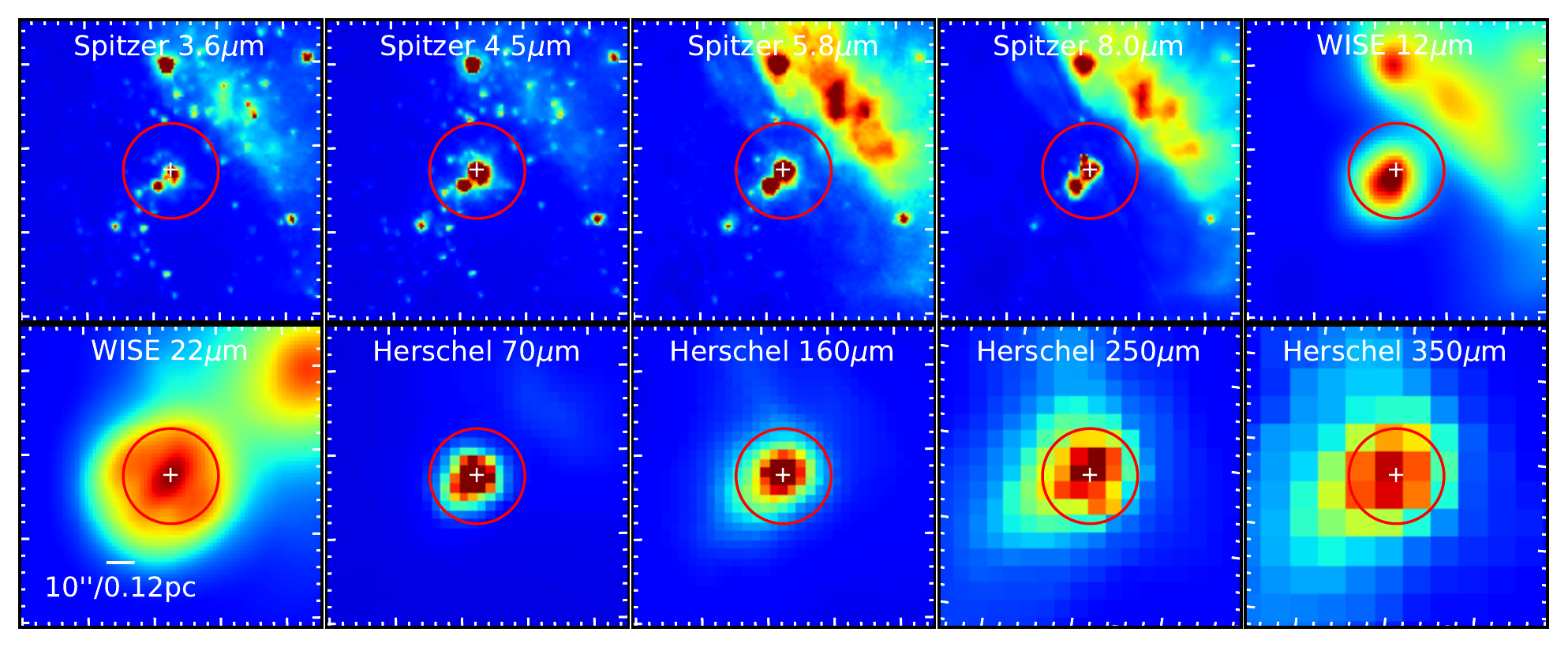}
\caption{Maps of G286c1 in different wavelengths observed with {\it Spitzer}, WISE and {\it Herschel}. The position of G286c1 is marked with a white cross. The red circles indicate the aperture used for photometry.\label{fig:map_sed}
}
\end{figure*}

To determine the properties of massive protostar(s) in the center of G286, we searched for infrared counterparts of the dense core G286c1 from near-IR to far-IR bands. We retrieved the datasets in the archive including {\it Spitzer} 3.5, 4.5, 5.8 and 8.0~$\mu$m, WISE~12, 22~$\mu$m and {\it Herschel} 70, 160, 250, 350, 500~$\mu$m maps. \autoref{fig:map_sed} presents a zoom-in view of the infrared images for each core from 3.6~$\mu$m to 350~$\mu$m. We use aperture photometry to measure the fluxes in each band, with a fixed aperture radius of 17\arcsec. This aperture size is determined using an automated method described in \citet{Fedriani22}, which is based on the gradient of the background-subtracted enclosed flux at 70~$\mu$m.


Aperture photometry is carried out following the method of \citet{Liu19}, i.e., we carry out a background subtraction using the median flux density in an annular region extending from one to two aperture radii, to remove general background and foreground contamination. The error bars are set to be the larger of either 10\% of the background-subtracted flux density or the value of the estimated background flux density. Note that the flux measurement for $\lambda<$8.0~$\mu$m is most likely overestimated since the adopted aperture also covers emission from surrounding sources. This will not significantly affect our SED fitting results since in our SED modeling the data points of shorter wavelength ($<$ 8.0~$\mu$m) are treated as upper limits (see \citet{Buizer17} for more details). We also note that we did not use the photometry at 500~$\mu$m in the SED fitting as the adopted aperture is smaller than the spatial resolution.

We utilize the python package {\it sedcreator} to fit the IR to millimeter SEDs towards G286c1 \citep{Fedriani22}, which is based on \citet{Zhang18} radiative transfer models (ZT models hereafter). The ZT model is a continuum radiative transfer model that describes the evolution of high- and intermediate-mass protostars with analytic and semi-analytic solutions based on the paradigm of the Turbulent Core model \citep[see][for more details]{Zhang18}. The main free parameters in this model are the initial mass of the core $M_c$, the mass surface density of the clump that the core is embedded in $\Sigma_{\rm cl}$, the protostellar mass $m_*$, as well as other parameters that characterize the observational setup, i.e., the viewing angle $\theta_\mathrm{view}$, and the level of foreground extinction $A_V$.

\autoref{fig:sed} shows the best fit SEDs for G286c1, with the parameters for the best five fitted models reported in \autoref{table:sed}. 
If we consider the typical parameter ranges among the best five models as an initial constraint for the protostellar system, G286c1 can be fitted with a protostellar source with a central mass of 4 -- 12~\msun, with an accretion rate of 0.9 -- 7.0 $\times 10^{-4}$~\msunyr. Such large ranges reflect the model degeneracy that exists in trying to constrain protostellar properties from only their MIR to FIR SEDs \citep[][]{Buizer17}, especially given that the fluxes at 10--50$\mu$m in not well constrained in our case. Following \citet{Fedriani22} we also list the parameters for the average and dispersion of the all the ``good'' models, defined as models that satisfy $\chi^2<$2. This averaged solution prefers a more massive protostar of $\sim$14.0$^{16.0}_{7.5}$~\msun, embedded in a relatively low-$\Sigma$ core 0.4$^{1.2}_{0.3}$~\gcm, and acrreting at 2.1$^{2.6}_{1.2}$ $\times 10^{-4}$~\msunyr.
A caveat in the SED analysis is that we have implicitly assumed the fluxes are mainly contributed by a single protostar, which may not be true for G286c1. G286c1 clearly hosts a triple system and the three components have comparable disk masses. 


\newpage
\begin{sidewaystable}
\startlongtable
\begin{deluxetable}{cccccccccccccc}
\tablecaption{Parameters of the best five fitted models for G286c1\tablenotemark{a,b}\label{table:sed}}
\tablehead{
\colhead{Source} & \colhead{$\chi^2$}  & \colhead{$M_c$} & \colhead{$\Sigma_{\rm cl}$} & \colhead{$m_*$} & \colhead{$R_{\rm core}$} & \colhead{$\theta_{\rm view}$} & \colhead{$A_V$} &  \colhead{$\theta_{\rm w,esc}$} & \colhead{$m_{\rm disk}$} & \colhead{$r_{\rm disk}$} &  \colhead{$\dot{m}_{\rm disk}$} & \colhead{$L_{\rm bol,iso}$} & \colhead{$L_{\rm bol}$} \\
\colhead{} & \colhead{}  & \colhead{$M_\odot$} & \colhead{$\rm g cm^{-2}$} & \colhead{$M_\odot$}& \colhead{pc} & \colhead{\arcdeg} & \colhead{mag} &  \colhead{\arcdeg} & \colhead{$M_\odot$} & \colhead{(AU)} &  \colhead{$M_\odot$/yr} & \colhead{$L_\odot$} & \colhead{$L_\odot$}
}
\startdata
 G286c1& 0.21 & 80 & 3.2 & 4.0 & 0.04 & 22 & 15.7 & 14 & 1.3 & 18 & 5.1 $\times10^{-4}$ & 0.9 $\times10^4$ & 0.8 $\times10^4$ \\
 d = 2.5~kpc& 0.28 & 80 & 3.2 & 8.0 & 0.04 & 22 & 230.8 & 19 & 2.7 & 30 & 7.0 $\times10^{-4}$ & 1.8 $\times10^4$ & 3.1 $\times10^4$ \\
 $R_{\rm ap}$ = 17\arcsec& 0.34 & 240 & 0.1 & 12.0 & 0.36 & 29 & 138.6 & 19 & 4.0 & 179 & 0.9 $\times10^{-4}$ & 2.0 $\times10^4$ & 1.7 $\times10^4$ \\
 $R_{\rm ap}$=0.21pc& 0.35 & 160 & 1.0 & 8.0 & 0.09 & 13 & 281.1 & 13 & 2.7 & 46 & 3.6 $\times10^{-4}$ & 1.3 $\times10^4$ & 6.2 $\times10^4$ \\
 & 0.36 & 480 & 0.3 & 8.0 & 0.29 & 13 & 168.3 & 8 & 2.7 & 65 & 2.0 $\times10^{-4}$ & 0.9 $\times10^4$ & 1.9 $\times10^4$ \\
{ Average model} & {(\#1325)} & 156$_{75}^{143}$ & 0.4$_{0.3}^{1.2}$ & 14.0$_{7.5}^{16.0}$ & 0.14$_{0.09}^{0.22}$ & 56$\pm 22$ & 194$\pm 149$ & 29$\pm 17$ & 4.7$_{2.5}^{5.3}$ & 116$_{71}^{187}$ & 2.1$_{1.2}^{2.6}$ $\times10^{-4}$ & 2.6$_{2.0}^{8.1}$ $\times10^4$ & 4.0$_{3.0}^{11.6}$ $\times10^4$ \\
\enddata
\end{deluxetable}
\tablenotetext{a}{The first five rows refer to the best five models taken from the 432 physical models, whereas the sixth row shows the average and dispersion of ``good'' model fits (see text). Upper and lower scripts in the average models row refer to the upper and lower end of the dispersion interval.}
\tablenotetext{b}{From left to right, the parameters are reduced $\chi^2$, the initial core mass $M_c$, the mean mass surface density of the clump $\Sigma_{\rm cl}$, the current protostellar mass $m_*$, the core radius $R_{\rm core}$, the viewing angle $\theta_{\rm view}$, foreground extinction $A_V$, half opening angle of the outflow cavity $\theta_{\rm w,esc}$, the mass of the disk $m_{\rm disk}$,  the radius of the disk $r_{\rm disk}$, accretion rate from the disk to the protostar $\dot{m}_{\rm disk}$, the luminosity integrated from the unextincted model SEDs assuming isotropic radiation $L_{\rm bol,iso}$, and the inclination-corrected true bolometric luminosity $L_{\rm bol}$. For the average model the second column refers to the number of ``good'' models.}

\end{sidewaystable}

\begin{figure*}[ht!]
\epsscale{0.8}\plotone{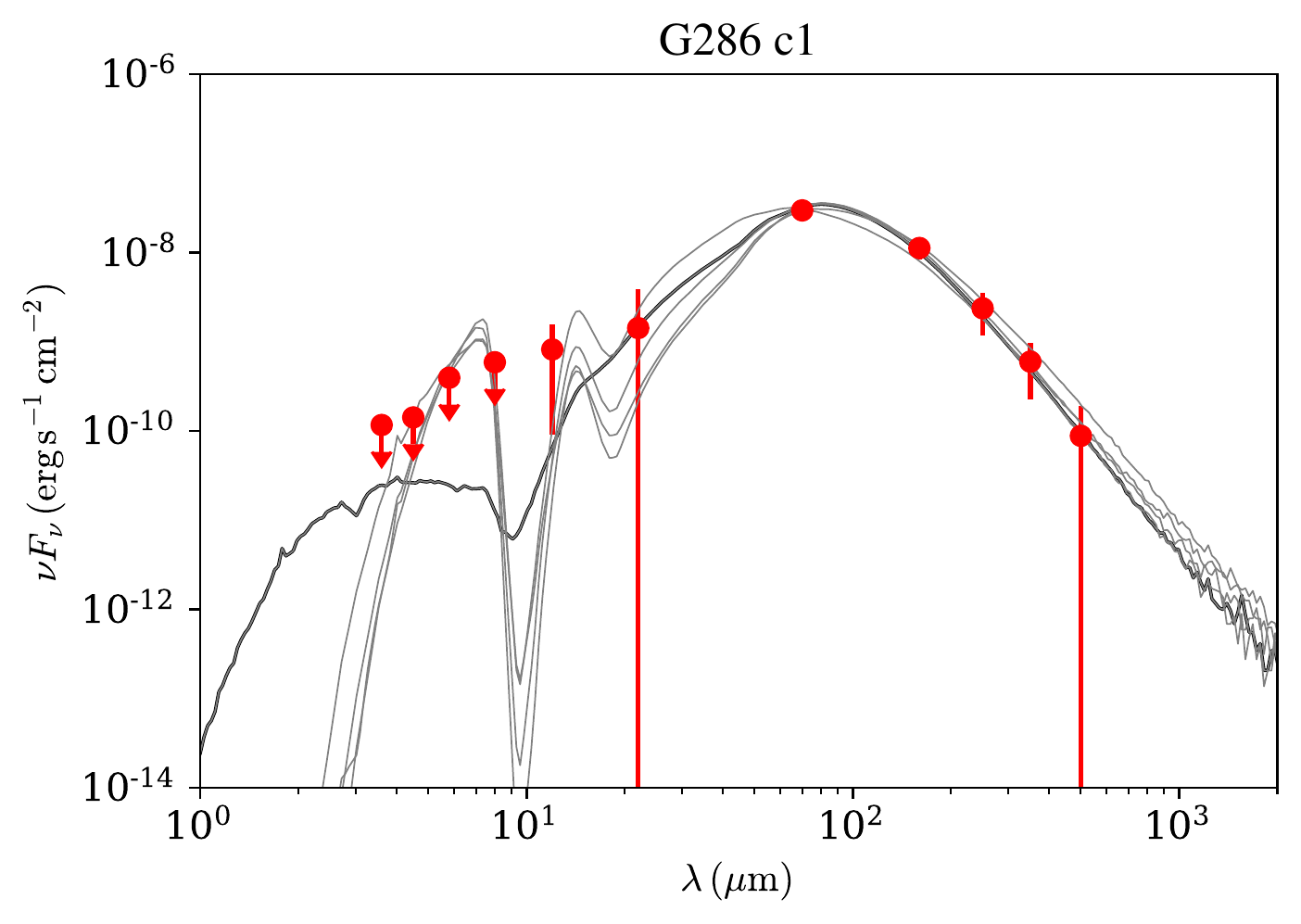}
\caption{Protostar model fitting to the fixed aperture, background-subtracted SED of G286c1 using the ZT model grid. The best-fit model is shown with a solid black line and the next four best models are shown with solid gray lines.\label{fig:sed}
\label{fig:sed}}
\end{figure*}


\acknowledgments
JCT acknowledges support from NSF grant AST-2009674 and ERC Advanced Grant 788829 (MSTAR). RF acknowledges funding from the European Union’s Horizon 2020 research and innovation programme under the Marie Sklodowska-Curie grant agreement No 101032092. JW acknowledges support by the National Science Foundation of China (NSFC) grant 12033004. This paper makes use of the following ALMA data: ADS/JAO.ALMA\#2017.1.00552.S. ALMA is a partnership of ESO (representing its member states), NSF (USA) and NINS (Japan), together with NRC (Canada), MOST and ASIAA (Taiwan), and KASI (Republic of Korea), in cooperation with the Republic of Chile. The Joint ALMA Observatory is operated by ESO, AUI/NRAO and NAOJ. The National Radio Astronomy Observatory is a facility of the National Science Foundation operated under cooperative agreement by Associated Universities, Inc.

%

\vspace{5mm}
\facilities{Atacama Large Millimiter/submillimeter Array (ALMA), Hubble Space Telescope (HST)}
\software{CASA \citep{McMullin07}, APLpy \citep{Robitaille12}, Astropy \citep{Astro13}}



\bibliography{refer}



\end{document}